# Generating grain boundary structures at a specified energy with a symmetry-constrained diffusion model

Kazuya Miyamoto[1,a)], Ibuki Okuda[1,b)], Izumi Takahara[1,c)], and Teruyasu Mizoguchi[1,2,a)]

[1]Department of Materials Engineering, the University of Tokyo, Tokyo, Japan

[2]Institute of Industrial Science, the University of Tokyo, Tokyo, Japan

[a)]Authors to whom correspondence should be addressed: miya1216@iis.u-tokyo.ac.jp and teru@iis.u-tokyo.ac.jp

[b)]Electronic mail: okuda@iis.u-tokyo.ac.jp

[c)]Electronic mail: KOUGEN@iis.u-tokyo.ac.jp

## Abstract

Grain boundaries (GBs) govern many properties of polycrystalline materials, and inverse design would make the GB structure itself a design variable. Crystal generative models such as MatterGen operate on periodic cells, which contain two GBs. The cell energy defines a single GB only when the two are equivalent, but existing models do not enforce this equivalence. Building on MatterGen, we developed a symmetry-constrained diffusion model in which the two GBs are related by a symmetry operation: atoms near one GB are generated, and those near the other are reconstructed through a symmetry projection applied at every denoising step. We applied it to the Σ5(210)[001] symmetric tilt GB in Al, conditioning generation on the GB energy from a machine-learning interatomic potential (MLIP). All 350 generated structures kept a geometrically equivalent GB pair after relaxation. Given a target GB energy, the model generated structures at that energy: for the three in-range targets with many nearby training structures, 44% to 70% fell within ±0.1 $J/m^2$ of the target. Tracking weakened at higher energies and failed outside the training range. Of the 91 near-target structures, 90 matched no training structure under the adopted geometric criterion, indicating that the model can propose configurations near the target energy that are distinct from the training structures, a capability that could help identify unknown metastable GB structures observed in experiments. Building GB-pair equivalence into generation thus makes GB properties well-defined design targets, enabling inverse design of GB structures.

## 1. Introduction

A grain boundary (GB) is an internal interface in polycrystalline materials[1]. Its structure and character are related to material properties such as solute segregation[2,3], GB migration, and fracture[4]. The GB energy is a basic quantity describing the thermodynamics of a GB; it depends on the misorientation and the GB plane, and also on the atomic arrangement near the interface[1,5]. Even for the same misorientation relationship and GB plane, several GB states with different atomic arrangements and densities may exist[5–9]. Given this structural diversity, we need to clarify how each atomic arrangement relates to the GB energy. This is important for understanding and controlling GB structure.

Conventional GB structure searches explore candidates by combining rigid in-plane translations with adjustments of the atomic arrangement and the number of atoms near the interface[10–12]. More recently, evolutionary algorithms and grand-canonical searches have been used to examine GB structures with different atomic densities systematically[13]. For example, Zhu et al. found multiple metastable GB phases of symmetric tilt GBs in Cu[6]. In addition, searches using machine-learning interatomic potentials (MLIPs) have been reported: Nishiyama et al. obtained GB structures and energies for face-centered cubic (FCC) metals including Al by local optimization of initial configurations with rigid-body translations[14], and similar searches exist for GBs in Si[15,16]. Interface informatics has also been applied, through virtual screening with machine-learned regression[17] and transfer learning[18]. These studies are useful for finding low-energy GB structures and multiple structural states.

Their search strategies, however, are directed toward energy minima[7], so metastable states are obtained only incidentally and the structures found depend on how the initial candidates are constructed. Yet metastable GB states can also form in real materials[5], and many distinct atomic arrangements may exist even at a single GB energy[19], some of them outside the range covered by a given forward search. Enlarging the search does not resolve this efficiently. The degrees of freedom of a GB (the rigid translations, the number of boundary atoms, and their arrangement) multiply, so the number of candidates grows combinatorially. MLIPs make each evaluation inexpensive but do not reduce this number, and a finite grid samples continuous variables only at its nodes. Moreover, a forward search learns the energy of a structure only after evaluating it, so obtaining structures at a specified energy means enumerating and discarding most candidates, increasingly so away from the most populated energy range, and anew for each target.

Inverse design instead obtains atomic structures for a target property directly. Given the target GB energy as a generation condition, structures, including metastable ones, can be sampled at specified energies without repeating the enumeration for each target. The two approaches are complementary: forward calculations supply the training data, and the generative model turns this finite set into a distribution from which structures can be drawn for new targets within its range. Because this distribution is over continuous variables rather than a fixed list of candidates, the model can also propose configurations between the nodes of the grid from which the training data were built.

In recent years, machine-learning generative models have been developed for such inverse design of materials[20,21]. For crystal structure generation, models that handle periodic atomic arrangements have been proposed, such as the Crystal Diffusion Variational Autoencoder (CDVAE) and the diffusion-based DiffCSP[22,23]. DiffCSP++, an extension of DiffCSP, builds the space group constraint of a crystal into the diffusion process[24]. For conditional generation, MatterGen has been proposed as a diffusion model[25] that generates the atomic species, coordinates, and lattice of inorganic crystals; by fine-tuning, it can generate structures conditioned on chemical composition, symmetry, and mechanical, electronic, and magnetic properties[26]. It has further been extended to jointly generate crystal structures and site-resolved local electronic descriptors (MatterGen-$e^{-}$)[27]. These studies build structural constraints or desired properties into the generation process. For bulk crystals, property-conditioned generation has thus become a practical alternative to enumeration.

The targets of such generative models are expanding from perfect periodic crystals to structures that include GBs and interfaces. Lei et al. developed a grand-canonical diffusion model that handles changes in the number of atoms and applied it to GB structures (tungsten twist GBs)[28]. Their model generates the GB region between fixed grains in a slab with vacuum, without a property condition, and evaluates the GB energies after relaxation; generation itself thus still follows the generate-then-evaluate sequence of a forward search. Conditioning a crystal generative model directly on the GB energy raises two GB-specific issues. The first is geometric validity: a GB is defined by two grains with a prescribed misorientation and boundary plane, and generating all atoms and the lattice without constraints guarantees neither this geometry nor the formation of the intended boundary. The second is energy assignment: a fully periodic cell, which avoids free surfaces, necessarily contains two GBs, and its excess energy represents a single GB only when the two are equivalent; otherwise it is an average and is ambiguous both as a training label and as a generation target. Existing crystal generative models address neither issue.

Here, we address both issues by extending MatterGen with a structural representation that links the two GBs by a symmetry operation and a symmetry projection that maintains this relation during diffusion. Fixing the grains to the prescribed geometry addresses geometric validity, and making the two GBs equivalent addresses energy assignment, so that the GB energy serves as a well-defined training label and generation target. Using Al $\Sigma5(210)[001]$ symmetric tilt GB as a model system, we evaluate the equivalence of the generated GB pairs against an unconstrained baseline, the tracking of target energies against unconditional generation and the training data, and the geometric distinctness of the generated structures from the training data.

## 2. Methods

### 2.1. Construction of the training data and grain boundary energy

To demonstrate the proposed method, we chose the $\Sigma5(210)[001]$ symmetric tilt GB in FCC Al as a model system. The training data were periodic bicrystals that differ in the relative translation between the grains, the rigid translation along the GB normal, and the number of atoms near the GBs. Under periodic boundary conditions, such a bicrystal cell necessarily contains two GBs; the two must be equivalent for the excess energy of the cell to represent a single GB (**Fig.** 1(a)). Initial structures were built with interface_master[29] by sampling the in-plane relative translation over the cell of non-identical displacement (CNID)[30]. Removing duplicates among these structures gave 31 templates (Section S1 of the supplementary material).

For each template, we moved one grain rigidly along the GB normal by $dx$ (10 levels), which changed the cell length along the normal by $2dx$. Next, we removed pairs of atoms related by a symmetry operation, one atom from each of the two GBs per pair, which keeps the symmetry of the periodic cell. This gave 6,510 structures. We relaxed the structures using an MLIP, MACE-MH-1[31,32] (*omat_pbe* head[33], double precision), and the FIRE method[34]. The cell lattice vectors and 76 bulk atoms were fixed, and only the atoms near the GBs were relaxed. All 6,510 structures converged.

After relaxation, we removed duplicates among structures with the same $dx$ and atom count using the pymatgen StructureMatcher, which left 5,509 structures. The data were split by template, giving 4,419 training, 538 validation, and 552 test structures.

We calculated the GB energy $\gamma$ from the excess energy of the periodic cell[35] as follows.

$$\gamma = (E_{\text{cell}} - NE_{\text{bulk}})/(2A). \tag{1}$$

Here, $E_{\text{cell}}$ is the total energy of the relaxed cell, $N$ the number of atoms, $E_{\text{bulk}}$ the energy per atom of bulk Al ($-3.757$ eV/atom), and $A = \|\boldsymbol{b} \times \boldsymbol{c}\|$ the area of one GB. The denominator is the combined area of the two boundaries, so $\gamma$ is the energy of one GB only when the two are equivalent (**Fig.** 1(a)); otherwise it generally represents their average. Because the cell and the bulk atoms are held fixed in this relaxation, $\gamma$ as used throughout this work is a value defined by this fixed procedure; the same definition serves as the training label, the generation condition, and the evaluation metric (Section S2).

### 2.2. Symmetry-constrained generative model

We built our symmetry-constrained GB generation model on MatterGen[26], a diffusion-based generative model for crystal structures. Our model relates the two GBs in a periodic cell through a symmetry operation (**Fig.** 1). We kept the denoising network of MatterGen (GemNet-T family[36]), its diffusion-based generation framework[37], and the Predictor-Corrector sampling[38], and added a structural representation adapted to GBs, generation variables, a symmetry projection, and associated update procedures (Section S3).

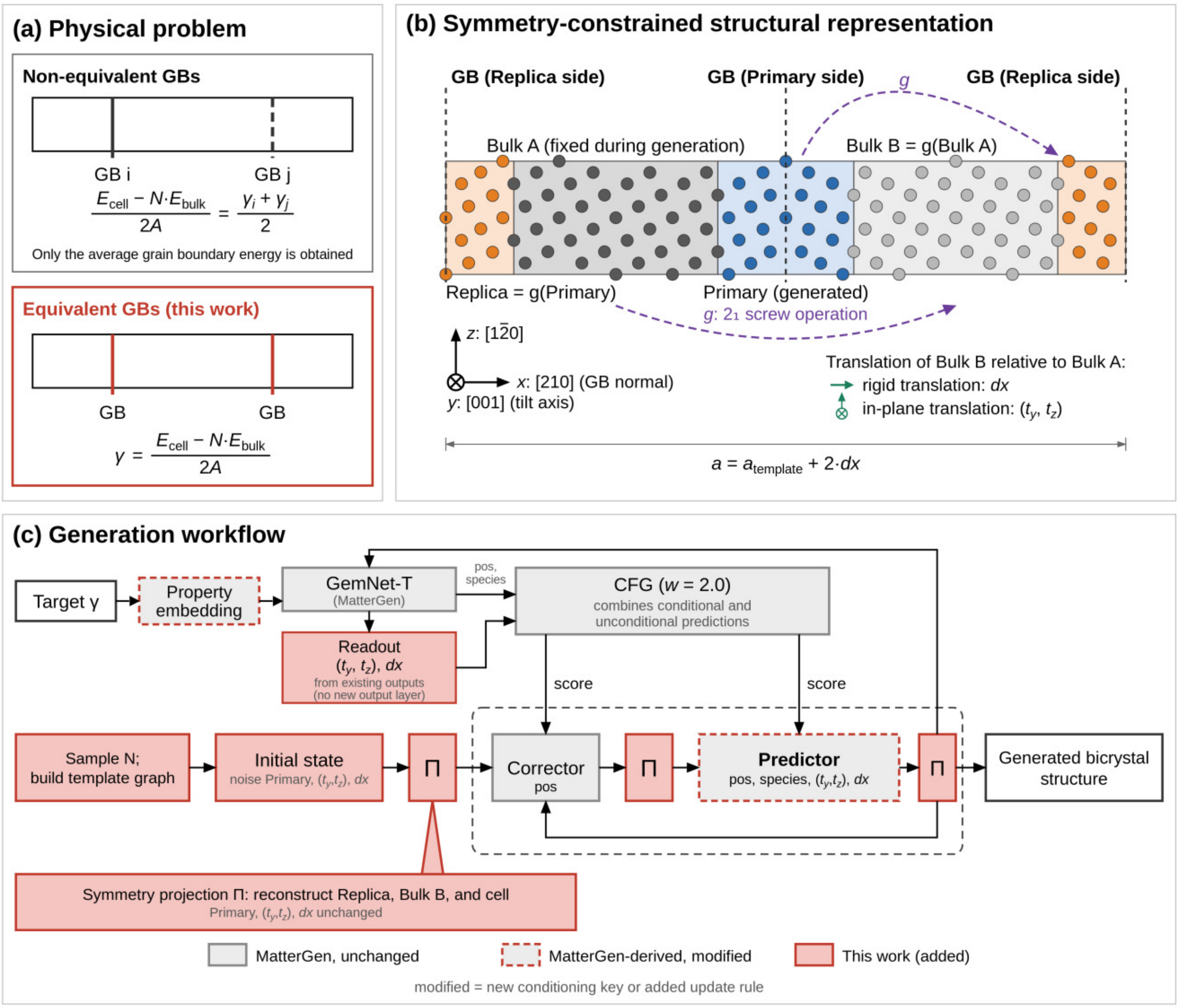

**Fig.** 1. Schematic of the constraint. (a) A periodic cell contains two GBs. When the two are non-equivalent, only their average GB energy is obtained from the excess energy of the cell. The solid and dashed lines distinguish the two GBs (GB i and GB j). (b) View along the tilt axis ($y$ ∥ [001], pointing into the page). The axis indicator gives the directions in Bulk A, $x$ = [210] (normal to the GB) and $z = [1\bar{2}0]$; in Bulk B, $x = [2\bar{1}0]$ and $z = [\bar{1}\,\bar{2}0]$. Dashed arrows denote the symmetry operation $g$. In the proposed model, the two GBs (the Primary side and the Replica side) are linked by the $2_1$ screw operation $g$. Only the atoms on the Primary side diffuse. The Replica side and Bulk B are reconstructed by the symmetry operation. During generation, only the 38 atoms of Bulk A are fixed. Blue is the Primary, orange is the Replica, dark gray is Bulk A, and light gray is Bulk B. The variables $dx$ and $(t_y, t_z)$ are the components of the translation of Bulk B relative to Bulk A along the normal and in the GB plane, respectively. (c) Generation workflow. The symmetry projection Π is applied to the initial state and after each Corrector and Predictor update. The dashed frame encloses the repeated Corrector, Π, Predictor, and Π steps. Gray boxes are unchanged from MatterGen, red dashed boxes are MatterGen-derived with a changed conditioning key or update rule, and red filled boxes were added in this work.

We partitioned the atoms into two bulk regions, Bulk A and Bulk B (38 atoms each), and the Primary and Replica groups near the GBs ($m$ atoms each); the total number of atoms is as follows.

$$N = 76 + 2m, \qquad m \in \{18,19,20,21,22\}. \tag{2}$$

The Primary atoms were generated independently; the Replica atoms and Bulk B were reconstructed from the Primary atoms and Bulk A by a symmetry operation (**Fig.** 1(b); Section S3).

To relate the two GBs, we used a $2_1$ screw operation $g$, a 180° rotation about the GB normal combined with a half-period translation, defined for fractional coordinates $\boldsymbol{f} = (f_x, f_y, f_z)$ and the in-plane relative translation $(t_y, t_z)$ as follows.

$$g(\boldsymbol{f}) = \left(f_x + 1/2,\ -f_y + 1/2 + t_y,\ -f_z + 1/2 + t_z\right) \bmod 1. \tag{3}$$

A Bulk B atom is obtained from the position $\boldsymbol{r}_A = (x_A, y_A, z_A)$ of the corresponding Bulk A atom as follows, with $dx$ the rigid displacement of Bulk B relative to Bulk A along the normal (Section S3).

$$\boldsymbol{r}_B = \left(x_A + a_{\text{template}}/2 + dx,\ -y_A + b/2 + t_y b,\ -z_A + c/2 + t_z c\right). \quad (4)$$

The independently generated variables were the fractional coordinates and elemental species of the Primary atoms, $(t_y, t_z)$, and $dx$; only the 38 Bulk A atoms were fixed during generation, unlike the 76 bulk atoms fixed in the later relaxation.

We introduced a symmetry projection $\Pi$ that enforces the prescribed symmetry at each generation step by reconstructing the Replica atoms, Bulk B, and the cell from the Primary atoms, Bulk A, $(t_y, t_z)$, and $dx$. During training, $\Pi$ was applied after noise addition. During generation, it was applied to the initial state and after each Corrector and Predictor update, in the order Corrector → $\Pi$ → Predictor → $\Pi$. Thus the structure supplied to the network after each update satisfied the prescribed symmetry (**Fig.** 1(c)).

The training loss was defined on the Primary atomic coordinates and species, the in-plane relative translation, and $dx$. The number of atoms $N$ was drawn before generation from the atom-count distribution of the training dataset, as in MatterGen, without conditioning on the target $\gamma$, and remained fixed during diffusion (Section S3).

This representation and projection made the structures immediately after generation invariant under the prescribed symmetry operation, to within single-precision floating-point accuracy, under four conditions (Section S3). However, this invariance alone does not guarantee the formation of GBs, the preservation of their equivalence after relaxation, or the mechanical or thermodynamic stability of the generated structures.

## 2.3. Conditioning on the grain boundary energy and the comparison model

We initialized the model with the pretrained MatterGen weights (*mattergen_base*), introduced $\gamma$ as a conditioning variable through the property adapter[26,39], and fine-tuned all parameters. Training and generation followed the default classifier-free guidance settings of MatterGen[40] (conditioning dropout of 20%; guidance weight $w = 2.0$, where $w = 0$ gives the unconditional prediction; Section S3).

For comparison, we fine-tuned MatterGen on the same dataset without the extensions of this study (hereafter, the unconstrained baseline), using the code base of this study with the extensions disabled. Both models share the pretrained weights, data split, training settings, and checkpoint selection (Table S2).

Despite these shared settings, the two models differed in their generation variables, structural representations, loss functions, sampling update procedures, and post-generation relaxation conditions. This comparison is therefore one of workflows and not an ablation that isolates the effect of the symmetry projection $\Pi$ alone.

## 2.4. Structure generation and relaxation

To evaluate target tracking and behavior outside the training range, we set seven target GB energies based on the range of 0.530 J/m$^2$ to 2.537 J/m$^2$ of the 4,419 training structures. Five targets lay inside and two outside this range (Section S5).

For each target, we generated 50 structures with each model (350 per model) in two runs of 25 structures each, with the same random seeds for both models; the structures of the two models do not necessarily correspond one-to-one. As a control, the same 350 structures were also generated with the guidance weight set to $w = 0$, under which the target value does not enter the prediction, using the same random seeds and checkpoint. To check reproducibility, we repeated the generation with the proposed model twice using different random seeds, giving three generation series including the generation described above (hereafter, the main experiment; Table S3).

Generated structures were relaxed with the same MLIP, FIRE method, and fixed cell as the training data. For the proposed model, the 76 Bulk A and Bulk B atoms were fixed and only the 36–44 Primary and Replica atoms were relaxed; for the unconstrained baseline, all atoms were relaxed. The symmetry projection was not applied during relaxation for either model.

The unconstrained baseline generated 20 structures with distorted cell angles, which were excluded before relaxation (Section S7). All remaining structures of both models converged within 500 steps. Fractions were evaluated with the requested count of 50 structures per target as the denominator, counting excluded structures as not meeting the condition.

## 2.5. Detection of grain boundaries and equivalence judgment

We evaluated the number of GBs and the equivalence of the two boundaries from the atomic arrangement alone, without the internal Primary–Replica labels, in three stages: detection (stage 0), whole-structure symmetry matching (stage 1), and local structural matching (stage 2) (Section S8).

In stage 0, we detected GBs from changes in the dominant local crystal orientation, estimated from nearest-neighbor bonds, between layers along the normal. A structure with no change was classified as *No orientation switch detected*; this does not establish the physical absence of a GB.

In stage 1, we assessed atomic correspondence across the entire structure under four candidate symmetry operations, each with a free three-component translation.

In stage 2, we matched the local window around each detected boundary to the other boundary in both directions and used the larger maximum deviation.

Atomic correspondence was determined with the Hungarian method[41]. For each candidate operation, we calculated the maximum distance between corresponding atoms, $d_{\mathrm{max}}$, after iterative translation refinement; the two GBs were classified as equivalent when the smallest $d_{\mathrm{max}}$ satisfied the following condition.

$$d_{\mathrm{max}} \leq 0.064\,\text{Å}. \tag{5}$$

Stages 1 and 2 were evaluated independently, and a structure contained an equivalent GB pair if either stage satisfied the criterion after relaxation; only stage 1 was applied before relaxation.

The threshold of 0.064 Å was set after inspecting the distribution of the maximum deviation in the main experiment (Section S8; sensitivity in Section S9). The criterion assesses geometric agreement and does not establish structural stability or strict crystallographic equivalence.

## 2.6. Tracking of the target and geometric comparison with training structures

We evaluated target tracking by the distribution of the relaxed GB energies $\gamma$ and by the near-target fraction, defined as the fraction of the 50 structures generated for each target whose relaxed $\gamma$ lies within ±0.1 J/m$^2$ of the target value $\gamma_{\mathrm{target}}$. We also counted how many of the 4,419 training structures fell within ±0.1 J/m$^2$ of each target.

We assessed whether each generated structure geometrically matched a relaxed training structure with the same number of atoms, using the StructureMatcher of pymatgen (version 2025.6.14) with *stol* = 0.025; a structure is called train-matched if it matched at least one training structure and train-unmatched otherwise. The matching settings, the all-atom root-mean-square (RMS) displacement to the nearest training structure, and the sensitivity to *stol* are described in Section S10.

# 3. Results

## 3.1. Generation of equivalent grain boundary pairs by the symmetry constraint

We evaluated whether the proposed model generated structures with two equivalent GBs, with the unconstrained baseline fine-tuned on the same dataset for comparison. Two GBs were detected in all 350 structures of the proposed model immediately after generation, and all were equivalent, with whole-structure $d_{\mathrm{max}}$ from $3.1 \times 10^{-6}$ Å to $6.9 \times 10^{-6}$ Å. The best-matching operation always included a 180° rotation about the GB normal with a half-period translation, so the prescribed $2_1$ screw symmetry was reproduced within numerical precision.

After relaxation with the MLIP, two GBs were still detected in all 350 structures, and all were equivalent (**Fig.** 2(a), (b)); the maximum $d_{\max}$ from whole-structure matching was 8.5 × $10^{-6}$ Å. This agreement reflects the symmetric starting structures and the equivariant potential under the specified relaxation conditions (Section 2.4) and does not guarantee equivalence under relaxation of all atoms or under asymmetric perturbations (Section S2 of the supplementary material).

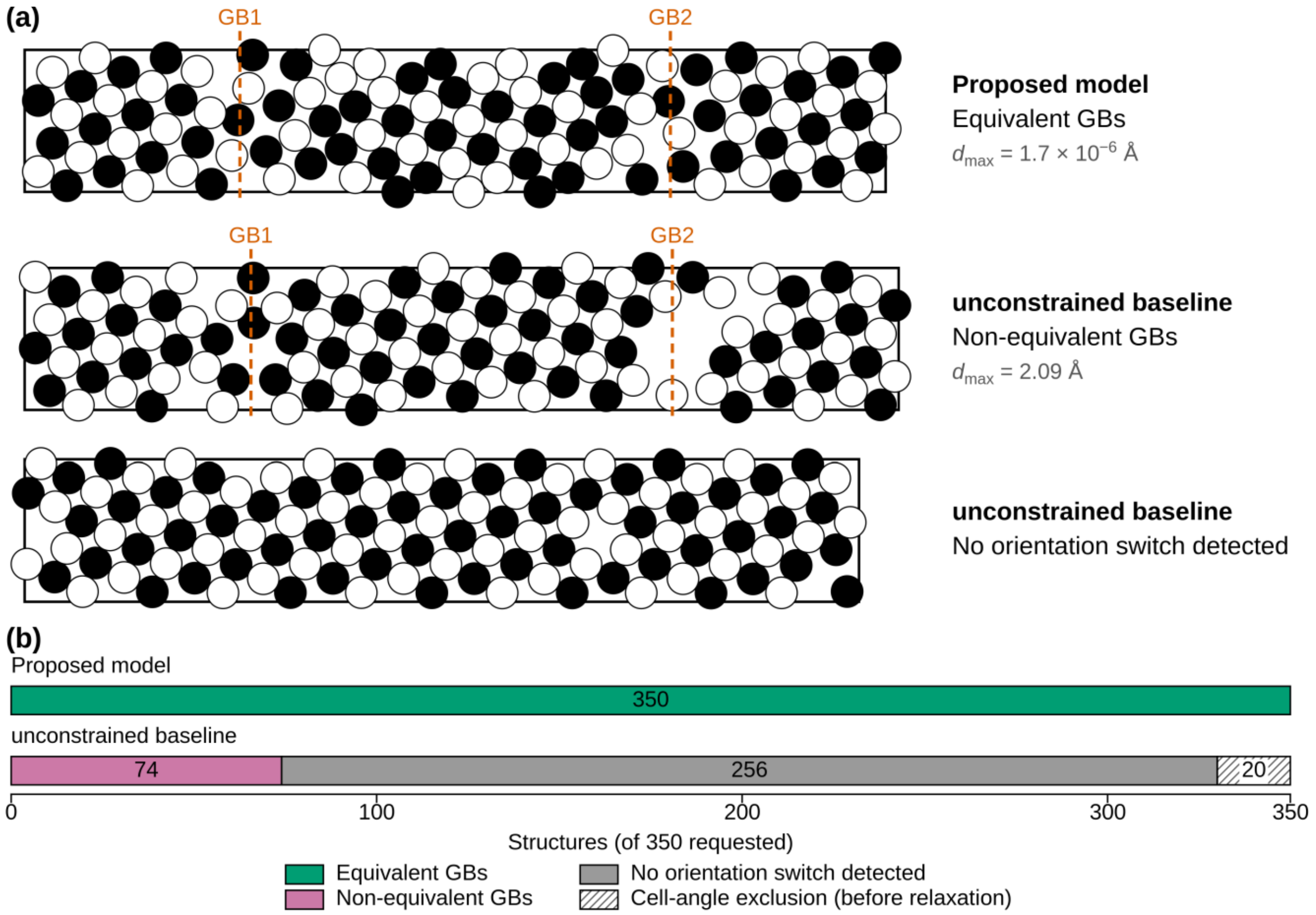

**Fig.** 2. Representative relaxed structures and class counts for the 350 structures requested from each model (Al, [001] projection). (a) From the top: a proposed-model structure with an equivalent GB pair, an unconstrained-baseline structure with a non-equivalent GB pair, and an unconstrained-baseline structure judged *No orientation switch detected*. Filled and open circles are the two atomic layers along the projection direction. Orange dashed lines mark the positions of the detected GBs (GB1 and GB2), and the black rectangle is the periodic cell. Structures with two detected GBs are shifted periodically to place the GBs symmetrically about the cell center. $d_{\max}$ is the stage 2 maximum deviation in the local window, and the equivalence threshold is 0.064 Å. Each structure was chosen by its own rule, so their targets differ. (b) All 350 proposed-model structures were equivalent. For the unconstrained baseline, 20 structures were excluded before relaxation because of cell-angle distortion (*Cell-angle exclusion*). Among the remaining 330 structures, none had an equivalent GB pair, 74 were non-equivalent, and 256 were *No orientation switch detected*. Relaxation conditions differ between the models (Section 2.4).

In contrast, 20 of the 350 structures requested from the unconstrained baseline were excluded before relaxation because of cell-angle distortion (Section 2.4). None of the remaining 330 relaxed structures contained an equivalent GB pair under the evaluation criteria of this study (**Fig.** 2(b)): two GBs were detected in 74 structures but were non-equivalent, and in the other 256 no layer-by-layer change in the dominant local orientation was detected (*No orientation switch detected*; **Fig.** 2(a)).

The proposed workflow produced equivalent GB pairs, so that the cell excess energy represents a single GB and can serve as a generation condition.

### 3.2. Structure generation conditioned on the grain boundary energy

The purpose of conditioning is not to find the lowest-energy structure, which forward searches already target, but to obtain structures at a specified energy, including metastable ones (Section 1). Conditioning is therefore assessed by how closely the relaxed energies follow the target, not by how low they are. Here, we generated 50 structures for each of five targets inside and two outside the training energy range (350 in total, all converged) and compared the relaxed $\gamma$ with the target by the distribution of $\gamma$ and the near-target fraction (**Fig.** 3, **Table** I). The generation with $w = 0$ from the same seeds for the target 1.132 J/m$^2$ is shown in Fig. 3(b). The unconstrained baseline was not evaluated here because it produced no equivalent GB pairs (Section 3.1).

**Table I.** Target tracking of the proposed model ($w$ = 2.0). For each target $\gamma$: the number of training structures within ±0.1 J/m$^2$ of the target, the mean and standard deviation (SD) of the relaxed $\gamma$ of the 50 generated structures, and the number of near-target structures with $w$ = 2.0 and with $w$ = 0 (same seeds). [a]Target outside the training range.

| Target $\gamma$ (J/m$^2$) | Training structures in window | Mean relaxed $\gamma$ (J/m$^2$) | SD (J/m$^2$) | Near-target ($w$ = 2.0) | Near-target ($w$ = 0) |
|---|---|---|---|---|---|
| 0.129[a] | 0 | 0.680 | 0.112 | 0 | 0 |
| 0.731 | 714 | 0.763 | 0.160 | 35 | 12 |
| 1.132 | 985 | 1.025 | 0.171 | 22 | 7 |
| 1.533 | 427 | 1.405 | 0.201 | 25 | 0 |
| 1.935 | 85 | 1.578 | 0.294 | 9 | 0 |
| 2.336 | 0 | 1.691 | 0.203 | 0 | 0 |
| 2.938[a] | 0 | 1.405 | 0.258 | 0 | 0 |

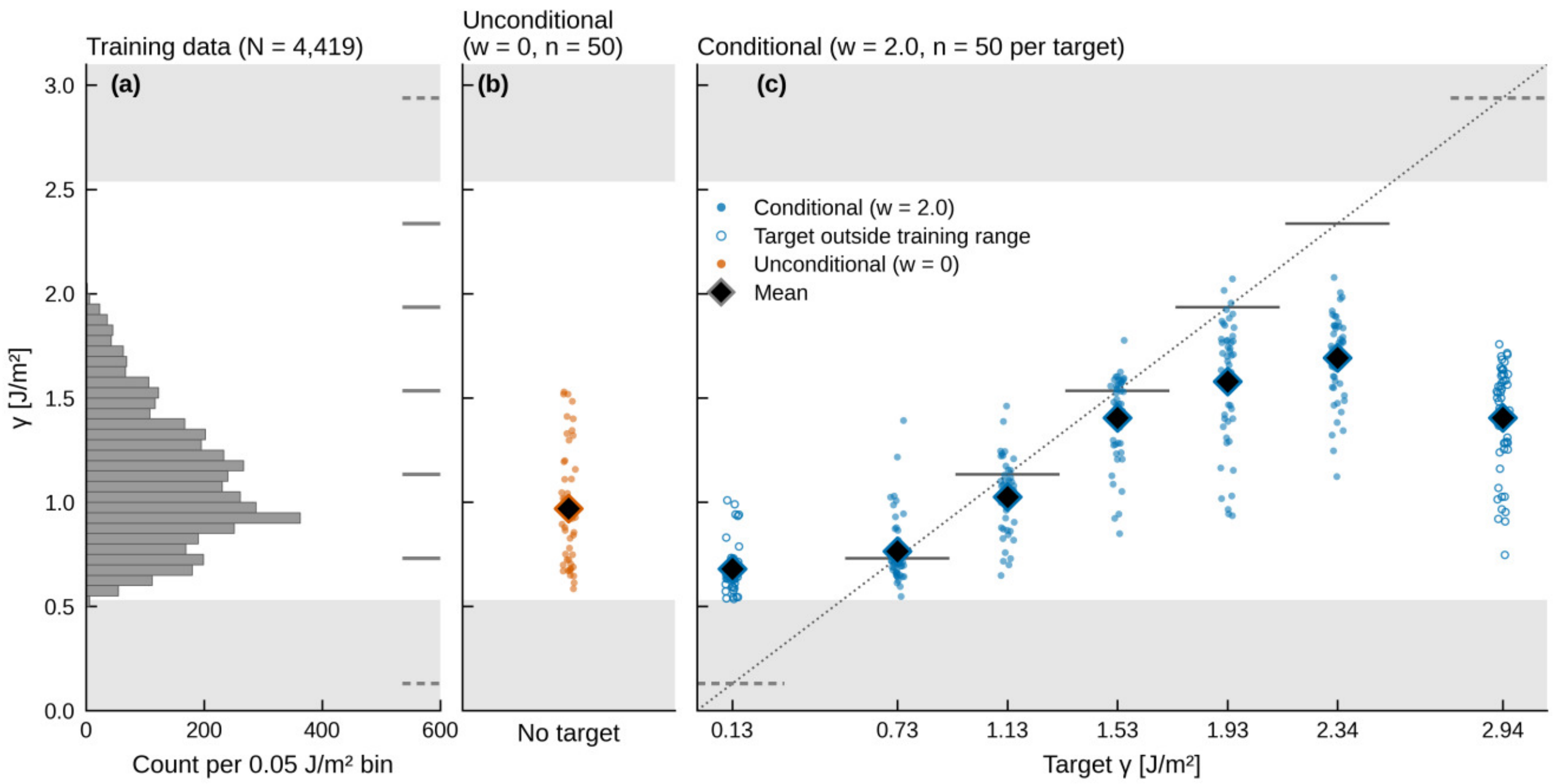

**Fig.** 3. Tracking of the target $\gamma$. The vertical axis is the $\gamma$ common to all panels. (a) Horizontal histogram of the $\gamma$ of the 4,419 training structures (bin width 0.05 J/m$^2$); short horizontal lines mark the seven target $\gamma$ values (dashed for the two targets outside the training range). (b) Relaxed $\gamma$ of the 50 structures generated for the target 1.132 J/m$^2$ with the guidance weight set to $w$ = 0, for which the target value does not enter the prediction, from the same random seeds and checkpoint as the main generation; the diamond marks the mean (0.969 J/m$^2$). (c) Relaxed $\gamma$ versus target $\gamma$ for the 50 structures generated at each target with $w$ = 2.0; black diamonds mark the means, gray horizontal lines mark the targets, the dotted line is $y = x$, and the light gray background marks the outside of the training $\gamma$ range. The two targets outside the training range are drawn with dashed lines and open circles. The number of structures that did not converge or were excluded is 0, and all 350 structures are plotted in panel (c).

At the targets of 0.731 J/m$^2$, 1.132 J/m$^2$, and 1.533 J/m$^2$, the mean relaxed $\gamma$ increased with the target (**Table** I), and the near-target fractions were 70%, 44%, and 50%, respectively. Across three generation series with different seeds, the near-target fractions ranged from 44% to 70% (Table S6). However, the standard deviations of $\gamma$ (0.160 J/m$^2$, 0.171 J/m$^2$, and 0.201 J/m$^2$) all exceeded the ±0.1 J/m$^2$ window, indicating that the energies were not tightly concentrated around their targets. Nevertheless, the conditioning produced, for a specified energy, a set of distinct structures rather than a single minimum.

At the higher targets 1.935 J/m$^2$ and 2.336 J/m$^2$, both within the overall training energy range, the near-target counts fell to 9/50 and 0/50, with 85 and 0 training structures within the ±0.1 J/m$^2$ window (**Table** I).

For the two targets outside the training energy range, 0.129 J/m$^2$ and 2.938 J/m$^2$, no generated structure was near-target (**Table** I). Tracking was therefore not shown outside the training range. Within the range, tracking failed at targets in energy regions sparsely represented in the training data. The factors that may limit tracking are discussed in Section 4.2.

To assess the effect of energy conditioning, we compared the near-target fractions of conditioned generation with those of unconditional generation ($w = 0$, one series) and of the training data. Unconditional generation yielded 12/50, 7/50, and 0/50 near-target structures at 0.731 J/m$^2$, 1.132 J/m$^2$, and 1.533 J/m$^2$ (Fig. S5), compared with 35/50, 22/50, and 25/50 with $w = 2.0$. At these three targets, the near-target fractions of the conditioned generation were 2.0 to 5.2 times those in the training data (Table S6). At 1.935 J/m$^2$, where training structures were sparse, the corresponding factor was at least 9 in the three series, although it rests on few near-target structures.

### 3.3. Geometric relation between the generated structures and the training data

Among the 250 structures generated with $w = 2.0$ for the five targets within the training energy range, 91 were near-target, and 43 of these had generated rigid translation values $dx$ along the GB normal outside the range specified during training. Of the 91 near-target structures, 90 did not match any relaxed training structure with the same number of atoms under the adopted criterion (StructureMatcher, *stol* = 0.025; Fig. S3). The number of train-unmatched structures depended on the tolerance, and the non-matches were not explained by $dx$ alone (Section S10). The minimum all-atom RMS displacement of the 90 train-unmatched structures to the nearest training structure had a median of 0.165 Å (range 0.050 Å to 0.573 Å). A train-unmatched classification does not establish the formation of a new GB phase, a new structural unit, or a new stable structure.

## 4. Discussion

### 4.1. Role of the symmetry constraint in conditional generation of grain boundaries

The proposed representation addresses the two issues identified in Section 1 (geometric validity and energy assignment) by fixing Bulk A and reconstructing Bulk B through the operation $g$, which keeps the prescribed grain geometry, and by reconstructing the Replica from the Primary through the symmetry projection $\Pi$, which makes the two GBs equivalent.

For the unconstrained baseline, mainly the first issue (geometric validity) remained: in 256 of the 330 relaxed structures no orientation switch was detected, and none of the 74 structures with two detected boundaries had an equivalent pair (Section 3.1). In the proposed model both issues are resolved by construction, and the 350/350 result shows that this was retained under the specified relaxation. This result itself is expected; the main contribution is that the constraint is compatible with energy conditioning: a network trained and sampled under $\Pi$ still produced structures whose energies followed the target (Section 3.2). We regard this compatibility, rather than the count itself, as the main contribution of the method. The equivalence is, however, not unconditional. Under asymmetric perturbations of 0.2 Å, 24 to 52 of the 350 structures relaxed to non-equivalent pairs with the bulk atoms fixed, whereas 349 to 350 structures remained equivalent under perturbations of 0.05 Å or less (Table S1 in the supplementary material). This suggests that for these structures the symmetric configuration is stable only against small perturbations: it either lies in a narrow basin of the unconstrained energy surface or is not a local minimum of it.

Because $g$ maps any arrangement at the Primary boundary to a valid image at the Replica, the constraint does not restrict the structure of an individual boundary; it excludes only cells with two different boundaries, whose excess energy cannot be assigned to one boundary anyway. It also halves the number of independently generated near-boundary atoms. Because the comparison was made between workflows (Section 2.3), the present results do not separate the contribution of $\Pi$ from that of the fixed-grain representation; a model with fixed grains but independently generated boundaries would make this separation.

### 4.2. Tracking and the energy distribution of the training data

Conditioning shifted the distribution of relaxed GB energies with the target, but the energies were pulled toward the range where training structures were dense and were not tightly concentrated at the specified value.

Near-target fractions were generally higher where more training structures lay near the target, but the relationship was not monotonic: the fraction at 1.132 J/m$^2$ was only 44% although its ±0.1 J/m$^2$ window contained the most training structures. Because the targets with many nearby training structures also lie near the center of the training energy distribution, the effects of the local density and of a pull toward the center cannot be separated with the present targets.

Without a controlled comparison in which the energy distribution of the training data was varied, that distribution cannot be identified as the cause of the limited tracking; the guidance strength, the model architecture, the number of atoms sampled independently of the target, and relaxation after generation may also contribute. Outside the training range, generation requires extrapolation of the learned relation between the condition and the structures, and the present model should not be used there. Within the range, applicability should therefore be judged from the local distribution of training structures around each target, not only from the energy range of the training data, and sparsely represented regions indicate where additional training structures are needed.

Sets of structures generated at a specified energy are not readily obtained by searches directed toward energy minima. They make it possible to ask how other GB properties, such as solute segregation or mobility, vary among structures that share the same GB energy, and they can supply candidates for comparison with experimentally observed GB structures, such as those imaged by scanning transmission electron microscopy, which are not necessarily ground states.

## 4.3. Interpretation of the non-match with the training structures

The displacement of the near-target, train-unmatched structures from their nearest training structures (median 0.165 Å, Section 3.3) can be compared with two reference scales (Section S10). It is smaller than the median distance between training structures that differ by one step in $dx$ (0.455 Å) (81 of the 90 structures fall below it). It is comparable to the median distance of test-split structures from templates not used for training (0.184 Å). The generated structures are therefore neither copies of training structures nor far outside the training distribution; they lie between the grid points from which the training data were constructed. This is consistent with the expectation in the Introduction that a generative model can propose configurations between the nodes of a finite forward-search grid.

## 4.4. Applicability to other grain boundary systems

The structural representation assumes that the two GBs of a periodic cell can be related by a known symmetry operation. Another geometry would require defining the symmetry operation, the independently generated atomic regions, and the regions derived through the operation. The operation that exchanges the two grains follows from the bicrystal geometry, and for symmetric boundaries it can be derived systematically; the grain regions and the CNID can be constructed automatically with existing tools for coincidence site lattice interfaces[29,30]. Extension to another boundary is therefore a matter of applying the same procedure rather than redesigning the model.

The main evaluation in this study used only the Al Σ5(210)[001] symmetric tilt GB. Applying the same model and training settings, with only system-specific constants changed, to the body-centered cubic (BCC) Fe Σ5(310)[001] symmetric tilt GB gave Spearman correlations of 0.808 and 0.766 between the target and MLIP energies before and after relaxation, with 43 of the 250 relaxed structures within ±0.1 J/m$^2$ of their targets; the baseline, density functional theory (DFT), and novelty comparisons were not repeated for Fe (Section S12). Other Σ values and tilt axes, asymmetric tilt boundaries, and twist boundaries have not been examined. For multicomponent GBs containing solute atoms[3,42], equivalence would have to account for both the symmetry of the atomic arrangement and the species occupying corresponding sites. In its present form, the model is fine-tuned for each boundary. Moreover, because the representation of fixed grains, a generated near-boundary region, and a symmetry projection is common to all boundaries, the macroscopic geometry of the boundary, such as the misorientation and boundary plane, can in principle be added as a further condition. A single model trained across many boundaries would then generate structures for boundaries and energies not seen individually during training.

### 4.5. Limitations on the energy evaluation and structural stability

The training labels, targets, and evaluation metrics in this study were all MLIP GB energies. As an auxiliary check, we performed DFT single-point calculations on 30 MLIP-relaxed structures drawn at random, stratified by target, from the 91 near-target structures (Fig. S8). The DFT energies were systematically lower than the MLIP energies, with a median DFT − MLIP difference of −0.128 J/m$^2$ that exceeds the ±0.1 J/m$^2$ window, so agreement with the MLIP target energies cannot be interpreted as agreement with absolute DFT energies. The Spearman correlation between the DFT and MLIP energies was 0.988 (95% bootstrap interval 0.955 to 0.996), and after subtracting the common median offset the mean absolute error was 0.041 J/m$^2$, well below the ±0.1 J/m$^2$ window. The relative ordering of energies on which the conditioning relies is therefore preserved. No DFT structural relaxation was performed, so these calculations do not establish structural stability. The generative model is independent of how the labels are computed; they can be replaced by DFT energies or by those of a more accurate potential.

The GB energies reported here are values of a fixed procedure in which the cell and the 76 bulk atoms were held fixed, under which all 350 generated structures satisfied the force criterion. When all atoms were allowed to relax, the energies changed by a median of −0.106 J/m$^2$, with their ordering largely preserved (Spearman correlation 0.959, Section S2). Vibrational analyses and finite-temperature evaluations were not performed, so thermodynamic metastability has not been established. The generated $dx$ fell outside the training range (−0.75 Å to +0.60 Å) for 207 of the 350 structures, whose physical validity has not been examined (Section S14).

## 5. Conclusion

Inverse design of GBs requires that the two boundaries in a periodic cell be equivalent, because only then does the cell excess energy represent a single boundary. We extended MatterGen with a Primary–Replica representation related by a $2_1$ screw operation and a symmetry projection in the diffusion process, so that the two GBs of a periodic bicrystal remain equivalent under energy conditioning. For the Al Σ5(210)[001] symmetric tilt GB, none of the structures from the unconstrained baseline had an equivalent GB pair under the evaluation criteria of this study, whereas with the symmetry projection all 350 generated structures retained an equivalent GB pair after relaxation with an MLIP. The GB energy thereby becomes a well-defined label for training and a well-defined target for generation.

The generated energies varied with the target: 44% to 70% of the structures fell within ±0.1 J/m$^2$ of the target at the three in-range targets with relatively many nearby training structures, fewer at higher targets, and none outside the training range. Relative to drawing structures from the forward-search data used for training, conditioning enriched the near-target fraction by factors of 2.0 to 5.2 at these three targets and of at least 9 at 1.935 J/m$^2$, where training structures were sparse. Of the 91 near-target structures, 90 matched no training structure under the adopted StructureMatcher criterion (median minimum all-atom RMS displacement 0.165 Å). This distance is smaller than one step of the training grid and comparable to that of structures from templates unseen during training; the model thus generated structures between the nodes of the training grid. Whether such differences correspond to new GB phases or structural units remains to be examined.

These findings support the feasibility of combining symmetry constraints with energy conditioning for the inverse design of periodic structures with an equivalent GB pair, but the present study is a proof of concept whose main evaluation is limited to a single Al symmetric tilt GB system, with a reduced check on BCC Fe, and to MLIP energies; structural stability and tracking of arbitrary target energies have not been established. The framework itself, however, is not specific to this boundary. The macroscopic geometry of the boundary and properties other than the energy, such as excess volume and solute segregation, could serve as conditions. Future work should examine structural stability, other GB geometries, and multicomponent systems. The present work establishes the first step toward these goals: a GB energy that can be specified, and structures that can be generated for it.

## Supplementary Material

See the supplementary material for the construction of the training dataset; the representation, symmetry projection, and guidance; the procedures for GB detection, equivalence assessment, and novelty evaluation; additional generation experiments; the application to BCC Fe Σ5(310)[001]; and the DFT single-point validation (Sections S1–S14), which includes additional references[43–49].

## Acknowledgments

This study was supported by the Ministry of Education, Culture, Sports, Science and Technology (MEXT) Supporting Pioneering Research through AI for 1,000 Discovery challenges Program (SPReAD), Japan, Grant Number JPMXP1726265519, Japan Science and Technology Agency (JST) (Nos. JPMJAX24DB and JPMJBS2418), and JSPS KAKENHI Grant Numbers JP25K24645, JP26K01205, and JP26K22608. Some computations were carried out using the computer resource offered by Research Institute for Information Technology, Kyushu University.

## Author Declarations

### Conflict of Interest

The authors have no conflicts to disclose.

### Author Contributions

**Kazuya Miyamoto:** Conceptualization (equal); Data curation (equal); Formal analysis (equal); Investigation (equal); Methodology (equal); Software (equal); Visualization (equal); Writing – original draft (equal). **Ibuki Okuda:** Software (equal). **Izumi Takahara:** Supervision (equal). **Teruyasu Mizoguchi:** Funding acquisition (equal); Supervision (equal); Writing – review & editing (equal).

## Data Availability

The data that support the findings of this study are openly available in Zenodo at https://doi.org/10.5281/zenodo.23667782. The deposit comprises the fine-tuned weights of the proposed model and the unconstrained baseline, the training dataset, and the generated, relaxed, and DFT-verified structures, together with the fine-tuned weights, training dataset, and generated and relaxed structures for the BCC Fe check. The code, including the symmetry-constrained extension of MatterGen and the evaluation and plotting scripts, is available at https://github.com/nmdl-mizo/mattergen-interface. The BCC Fe check used an adapted copy of the same code with iron-specific constants; its resolved training configuration is part of the Zenodo deposit. VASP pseudopotential files are not redistributed because of their license.

# Supplementary Material

## S1. Construction of the training dataset

The training structures are periodic bicrystals of the Σ5(210)[001] symmetric tilt grain boundary (GB) in face-centered cubic (FCC) Al. They differ in the relative translation between the grains, the rigid translation of a grain along the GB normal, and the number of atoms near the GBs. Initial structures were built with interface_master[29]. In-plane relative translations were sampled on a 10 × 10 grid of 100 points. The grid divides each of the two basis vectors of the cell of non-identical displacement (CNID)[30] into 10 equal intervals. The resulting orthorhombic cells contained 120 atoms. The cell length $a$ was 54.156 Å, $b$ was 4.037 Å, and $c$ was 9.026 Å, with the $a$ axis along the GB normal. We compared all 4,950 pairs among the 100 structures using the StructureMatcher of pymatgen (version 2025.6.14)[43]. The settings were *stol* = 0.025, *ltol* = 0.2, and *angle_tol* = 5°, with *primitive_cell*, *scale*, and *attempt_supercell* disabled. Structures that matched one another were grouped together, directly or through a chain of matches. If A matched B and B matched C, then A and C were placed in the same group. This gave 31 groups. We selected one structure from each group to make 31 templates. For each template, we moved one grain rigidly along the GB normal. The displacement $dx$ took 10 values, from −0.75 Å to +0.60 Å in steps of 0.15 Å. The other grain was not moved, and the cell length along the normal was changed by $2dx$, as follows.

$$a = a_{\text{template}} + 2dx. \tag{S1}$$

Here, $a_{\text{template}}$ is 54.156 Å. This operation changed the spacing between the grains by $dx$ at each of the two GBs in the periodic cell. The in-plane cell lengths and the atomic arrangement inside each grain were left unchanged. Next, we removed pairs of atoms related by a symmetry operation from the neighborhoods of the two GBs, one atom per GB for each pair. The number of removed pairs $k$ was varied over 0–4. When no pair was removed, there was a single pattern. For each $k$ with removal, we made five removal patterns. Removing the paired atoms from both GBs keeps the symmetry of the periodic cell after removal. The number of atoms is $N = 120 - 2k$, taking the five values 112, 114, 116, 118, 120. From 31 templates, 10 values of $dx$, and 21 removal patterns, we obtained 6,510 structures in total. The structures were relaxed with the machine-learning interatomic potential (MLIP) and the FIRE method described in Section 2.1, with a maximum force of 0.02 eV/Å and at most 500 steps. After relaxation, we used the StructureMatcher to compare all pairs of structures with the same $dx$ and atom count, including structures obtained from different removal patterns of the same template. We used the same software version and matching settings as described above. Structures that matched one another, directly or through a chain of matches, were placed in the same group, and one structure was retained from each group. This removed 1,001 structures and left a dataset of 5,509 structures. We then compared all pairs of the retained structures and confirmed that no duplicates remained under the same matching criterion. The 31 templates were split into 25 training, 3 validation, and 3 test templates, so that structures derived from the same template were not distributed across different subsets. For $E_{\text{bulk}}$ of the GB energy defined in Section 2.1, we used a 4-atom FCC Al cell whose lattice constant (4.037 Å) was optimized with the MLIP. The test split, which consists of templates not used for training, is not used for model selection; it serves only as a reference scale for geometric distances to the training data.

## S2. Robustness of the equivalence and of the fixed-procedure grain boundary energy

The equivalence of the two GBs in the relaxed structures of the proposed model follows from a symmetric starting structure and an equivariant potential, so we examined whether it survives asymmetric perturbations of the starting positions and the release of the fixed bulk atoms, using 20 perturbation conditions applied to all 350 structures of the main experiment (in this supplementary material, the generation experiments reported in the main text). The perturbations displaced selected atoms by 0.02 Å, 0.05 Å, 0.1 Å, or 0.2 Å, and relaxation was carried out either with the bulk atoms fixed as in the main experiment or with all atoms free; all relaxations converged. The perturbed atoms were the Primary and Replica atoms closer to one of the two GBs (perturbing only one boundary), all Primary and Replica atoms, or all atoms; when all atoms were perturbed, only the relaxation with all atoms free was carried out. The numbers of structures classified as containing an equivalent

GB pair ($d_{\max} \leq 0.064$ Å) are listed in **Table S**1; without perturbation, all 350 structures remained equivalent with and without the constraint. In the fixed-bulk relaxation of the main experiment, the residual force on the fixed atoms had a median of 0.011 eV/Å and a maximum of 1.11 eV/Å, 40% of the fixed atoms exceeded the convergence criterion of 0.02 eV/Å, and 77% of the squared force was carried by the component normal to the boundary, concentrated near the interface with the relaxed region. When all atoms were allowed to relax from the main-experiment structures, the GB energy changed by a median of $-0.106$ J/m$^2$, and $|\Delta\gamma|$ exceeded 0.05 J/m$^2$ for 251 of the 350 structures; the ordering of the energies was nevertheless preserved (Spearman correlation $\rho$ = 0.959, 95% bootstrap interval 0.939 to 0.972; within-target values 0.720 to 0.894). The same shift occurs for the training data: for the 2,752 training structures whose labels lie within ±0.1 J/m$^2$ of one of the targets, full relaxation changed $\gamma$ by a median of $-0.105$ J/m$^2$, with $\rho$ = 0.874 between the fixed-procedure and fully relaxed values. The GB energies reported in the main text are therefore values of the fixed procedure, and they shift under full relaxation by an amount comparable to the ±0.1 J/m$^2$ window. Finally, the shortest interatomic distance in the generated structures ranged from 2.417 Å to 2.797 Å (median 2.630 Å), within the range of the training structures (2.350 Å to 2.837 Å, median 2.596 Å).

**Table S**1. Number of structures (of 350) classified as containing an equivalent GB pair after relaxation, by perturbation amplitude and relaxation constraint. Ranges span the perturbation types examined at each amplitude.

| Perturbation amplitude (Å) | Bulk atoms fixed | All atoms free |
|---|---|---|
| None | 350 | 350 |
| 0.02 | 350 | 349–350 |
| 0.05 | 349 | 348–349 |
| 0.1 | 333–344 | 342–347 |
| 0.2 | 298–326 | 318–335 |

## S3. Representation, symmetry projection, and guidance

Predictor-Corrector sampling is an iteration that removes noise step by step from a noisy initial state. At each step, the Predictor moves the state to the next, less noisy step using the output of the denoising network. The Corrector then refines the state at the same step using the same output. In this study, the Corrector updates only the atomic coordinates, and the Predictor updates the atomic coordinates, elemental species, in-plane relative translation, and $dx$ (Fig. 1(c)). We defined the $x$ axis along the GB normal, the $y$ axis parallel to the [001] tilt axis, and the $z$ axis within the GB plane and perpendicular to both. In the operation $g$ of Eq. (3), $t_y$ and $t_z$ are dimensionless variables describing the in-plane relative translation between the grains, and their corresponding real-space displacements are $t_y b$ and $t_z c$, respectively. Under periodic boundary conditions, $g^2$ is the identity operation regardless of the values of $(t_y, t_z)$. The position of an atom in Bulk B is obtained from its position in Bulk A by Eq. (4), and the coordinates are wrapped into the periodic cell. Here, $dx$ represents the rigid displacement of Bulk B relative to Bulk A along the GB normal, rather than the change in cell length itself. The Replica atoms were reconstructed from the Primary using the same operation, with identical elemental species assigned to corresponding atoms. During generation, the Cartesian coordinates of the 38 Bulk A atoms were held fixed, and Bulk B, the Replica atoms, and the cell were reconstructed from the generated variables. The symmetry projection $\Pi$ updates the Replica atoms, Bulk B, and the cell according to the symmetry operation. The lattice was not an independent diffusion variable. It was reconstructed from $dx$ at each projection step. Predictions for the in-plane relative translation and $dx$ were obtained from existing network outputs. For the in-plane relative translation, we used the mean of the coordinate outputs of the Replica atoms. For $dx$, we used the (0,0) component of the output corresponding to the lattice. No new dedicated output layers were added. In each generation run, $N$ is first drawn from the atom-count distribution of the training dataset and the template graph for that $N$ is built. The initial state is then set by placing noise on the Primary coordinates, $(t_y, t_z)$, and $dx$ (Fig. 1(c)). No atoms were inserted or removed during diffusion. The invariance of the structures immediately after generation under the prescribed symmetry operation held only when four conditions were met. The Replica atoms were the images of the Primary atoms under $g$. Bulk B was the image of Bulk A under $g$. The elemental species of the Replica and the Primary atoms agreed, with Bulk A and Bulk B set to the parent element Al. The cell was orthorhombic. In the guidance equation below, $s$ represents the predicted score for the continuous variables (atomic coordinates, in-plane relative translation, and $dx$) and, for the elemental species, the logits that predict the denoised species. Guidance was applied to the predictions of the Primary atomic coordinates and species, the in-plane relative translation, and $dx$, and the guided predictions were used in both the Predictor and the Corrector steps.

During generation, we used classifier-free guidance (CFG)[40] (Fig. 1(c)), combining the conditional prediction $s_{\mathrm{cond}}$ and the unconditional prediction $s_{\mathrm{uncond}}$ as follows.

$$s_{\mathrm{guided}} = w\, s_{\mathrm{cond}} + (1 - w) s_{\mathrm{uncond}}. \tag{S2}$$

Here $w = 0$ and $w = 1$ give the unconditional and conditional predictions, and $w > 1$ extrapolates beyond the latter. This $w$ equals the guidance coefficient in the original formulation plus 1. We used $w = 2.0$ for the main generation experiments.

## S4. Training dataset and workflow comparison

The construction of the training dataset is described in Section S1; **Table S**2 compares the workflows of the proposed model and the unconstrained baseline.

**Table S**2. Comparison of the workflows of the proposed model and the unconstrained baseline, listing the settings that are the same and those that differ. D3PM, discrete denoising diffusion probabilistic model; VE, variance exploding; VP, variance preserving; SNR, signal-to-noise ratio.

| Category | Item | Proposed model | Unconstrained baseline |
|---|---|---|---|
| Same | Rows of the training data | train 4,419 structures, val 538 structures (the same rows of the same split) | Same as left |
| Same | Optimization settings | AdamW, learning rate $5 \times 10^{-6}$ with cosine decay to $10^{-7}$ over 200 epochs, batch size 4, gradient clipping by value at 0.5, 32-bit precision; validation every 5 epochs; the checkpoint with the lowest validation loss is used | Same as left |
| Same | Training seed and standardization of $\gamma$ | Seed 20290000; $\gamma$ standardized by the mean 1.116 J/m$^2$ and the standard deviation 0.311 J/m$^2$ of the training data | Same as left |
| Same | Pretraining | mattergen_base (revision 5244495dd9a9); 281 parameter tensors copied, 24 newly initialized | mattergen_base (revision 5244495dd9a9); the numbers of copied and new tensors were not logged |
| Same | Probability of dropping the condition | 20% per structure, for classifier-free guidance | Same as left |
| Same | Distribution of the number of atoms | Drawn from the distribution of the number of atoms in the training data | Same as left |
| Same | Target $\gamma$ | The same 7 targets | Same as left |
| Same | Relaxation software and settings | MACE-MH-1 (head omat_pbe), FIRE, fixed cell | Same as left |
| Different | Loss | Applied to the coordinates and species of the Primary, the in-plane translation, and the cell extension along the normal | Applied to the coordinates and species of all atoms and to the cell |
| Different | Loss weights | Coordinates 0.1, species 1.0, translation 1.0, cell extension 1.0, D3PM hybrid $\lambda$ 0.01 | Coordinates 0.1, species 1.0, cell 1.0, D3PM hybrid $\lambda$ 0.01 |
| Different | Noise schedules | Coordinates VE $\sigma_{\max}$ 5.0; translation $\sigma$ 0.01–5.0; cell extension VP $\beta$ 0.1–20; species D3PM mask, 1,000 steps | Coordinates VE $\sigma_{\max}$ 5.0; lattice VP $\beta$ 0.1–20; species D3PM mask, 1,000 steps |
| Different | Variables that diffuse | Coordinates and species of the Primary, the in-plane translation, and the cell extension along the normal | Coordinates and species of all atoms, and the cell |
| Different | Data representation | The relaxed structure ordered as the bulk on both sides, Primary, and Replica, with its actual cell | A crystal structure passed through the standard data transformations of MatterGen |
| Different | Best epoch | 159 | 129 (the loss is defined differently, so this is not the best on the sa me criterion) |
| Different | Initial state of generation | The bulk part and the cell are given by the construction; the Primary, the in-plane translation, and the cell extension are generated from noise | All atoms and the lattice are generated from noise |

| Category | Item | Proposed model | Unconstrained baseline |
|---|---|---|---|
| Different | Sampler | 1,000 steps, ancestral predictors (wrapped for coordinates and translation, D3PM for species), Langevin corrector on coordinates with SNR 0.4 and 1 step; guidance weight $w$ (Section S3) | The same, with an additional predictor and corrector for the lattice |
| Different | Atoms fixed during relaxation | 76 | 0 |
| Different | Exclusion before relaxation | 0 structures | 20 structures (cell-angle distortion) |

## S5. Additional generation experiments and reproducibility

With $W$ (2.007 J/m$^2$) the width of the training energy range, five targets were placed at 10%, 30%, 50%, 70%, 90% of the range from its lower bound (0.731 J/m$^2$, 1.132 J/m$^2$, 1.533 J/m$^2$, 1.935 J/m$^2$, and 2.336 J/m$^2$), and the extrapolation targets 0.129 J/m$^2$ and 2.938 J/m$^2$ each lay 0.2W beyond the corresponding bound. The runs used in this work are listed in Table S3. The main experiment is as defined in Section S2. In these experiments, both the proposed model and the unconstrained baseline were evaluated at $w = 2.0$ with seven target energies and 50 generated structures per target (350 structures per model). A regeneration is an additional run with the same target energies and number of structures as the main experiment. Only the initial random seed is changed. Regeneration probes only the variation associated with generation sampling. Each model was trained with one training seed. We also ran additional experiments at guidance strengths of $w = 1.0, 2.0, 3.0, 4.0$. Each used seven targets and 50 structures per target (350 structures per strength). The target energy set in the additional experiments differed from that of the main experiment. The additional experiments covered approximately 0.32 J/m$^2$ to 2.75 J/m$^2$, while the main experiment covered approximately 0.13 J/m$^2$ to 2.94 J/m$^2$.

In the two regenerations of the proposed model, 350 of 350 structures and 350 of 350 structures, respectively, contained equivalent GB pairs. The proposed model was regenerated only two times. The main experiment, the two regenerations, and the additional experiments at $w = 1.0, 2.0, 3.0, 4.0$ make up seven generation runs. In total, 2,450 structures were requested in these runs, excluding the $w = 0$ control. All 2,450 of the 2,450 structures were classified as containing equivalent GB pairs before relaxation. Among the 50 structures generated at $w = 4.0$ for the lowest target energy (0.321 J/m$^2$), 5 contained 2 to 4 atoms of elements other than Al. Because the symmetry projection copies the element types as well, these atoms were also paired as identical species between the two GBs. Because the Al GB energy used in this study could not be evaluated for these compositions, these 5 structures were excluded from relaxation and energy evaluation. All remaining 2,445 structures retained equivalent GB pairs after relaxation.

Across the four unconstrained-baseline runs (the main experiment and the three regenerations), 66 to 76 of the 350 structures in each run contained two GBs. Among those structures, the fraction classified as equivalent was at most 3.9%. The statement that the main-experiment conditions yielded very few equivalent GB pairs refers to these four runs. Of the 282 structures containing two GBs across the unconstrained-baseline main experiment and its three regenerations, 3 had grains too thin for the local structural matching (stage 2) of the equivalence evaluation (Sections 2.5 and S8) to be applied. In all 3 cases, the maximum deviation in the whole-structure symmetry matching (stage 1) exceeded the tolerance of 0.064 Å, and none was classified as equivalent at the first stage. In the additional experiments, only 40 to 68 of the 350 structures generated by the unconstrained baseline at each guidance strength contained two detectable GBs. At $w = 1.0, 2.0, 3.0, 4.0$, the numbers classified as containing equivalent GB pairs were 8, 1, 1, and 4, respectively. In the unconstrained-baseline main experiment, 20 structures were excluded because of cell-angle distortion. The corresponding counts for the additional experiments are given in Table S4. The intervals reported in Table S4 are Wilson 95% intervals[44] for the fraction of structures containing two GBs.

The GB energies of the generated structures at each guidance strength, before and after relaxation, are shown in Section S6.

**Table S3.** Runs used in this work. The seed column gives the base value of the random seed; the seed of each generation run is the base value plus the index of the run. Targets × structures gives the number of target GB energies and the number of structures per target. BCC, body-centered cubic; DFT, density functional theory.

| Run | Model | Guidance $w$ | Targets × structures | Seeds | Purpose | Section |
|---|---|---|---|---|---|---|
| Main experiment | Proposed | 2.0 | 7 × 50 | 20301000 | Equivalence, target tracking, and comparison with the training structures | Sections 3.1–3.3 |
| Main experiment | Unconstrained baseline | 2.0 | 7 × 50 | 20301000 | Equivalence of the baseline | Section 3.1 |
| Regeneration, series 2 | Proposed | 2.0 | 7 × 50 | 20311000 | Reproducibility of the near-target counts and of the equivalence | Sections S5, S11 |
| Regeneration, series 3 | Proposed | 2.0 | 7 × 50 | 20312000 | Reproducibility of the near-target counts and of the equivalence | Sections S5, S11 |
| Unconditional generation | Proposed | $w = 0$ | 7 × 50 | 20301000 (the seeds of the main ex periment) | Control for the guidance | Section S11 |
| Regenerations (3 runs) | Unconstrained baseline | 2.0 | 7 × 50 in each run | 20311000, 20312000, 20313000 | Reproducibility of the outcomes of the baseline | Sections S5, S9 |
| Additional experiments | Proposed and unconstrained baseline | 1.0, 2.0, 3.0, 4.0 | 7 × 50 at each strength (a different set of targets) | 20301000 | Dependence on the guidance strength | Sections S5, S6 |
| DFT single-point calculations | Proposed (MLIP-relaxed structures) | 2.0 | 30 structures drawn from the 91 near-target structures | not applicable | Check of the MLIP energies | Section S13 |
| Perturbation tests | Proposed | 2.0 | 350 structures × 20 conditions | not applicable | Robustness of the equivalence | Section S2 |
| Full relaxation | Proposed | 2.0 | 350 structures | not applicable | Fixed-procedure energy versus full relaxation | Section S2 |
| Full relaxation of the training data | Training structures | not applicable | 2,752 structures | not applicable | The same comparison for the training labels | Section S2 |
| BCC Fe check | Proposed, fine-tuned for BCC Fe Σ5(310)[001] | 2.0 | 5 × 50 (targets inside the training range) | 20301000 | Application of the procedure to a BCC boundary | Section S12 |
| MLIP check for Fe | BCC Fe GB structures selected by MLIP energy before generation | not applicable | 12 structures | not applicable | MLIP versus spin-polarized DFT energies for Fe | Section S12 |

**Table S4.** Outcomes of the additional experiments at each guidance weight $w$: for the unconstrained baseline, the number and fraction of structures containing two GBs with Wilson 95% intervals[44], the fraction of those classified as equivalent, and the numbers of structures classified as *No orientation switch* detected, with cell-angle distortion (excluded), with an unidentified axis, and with thin grains; and the number of equivalent structures for the proposed model.

| Guidance $w$ | Two GBs (count, fraction, interval) | Equivalent (fraction of two-GB structures) | No orientation switch detected | Cell-angle exclusion | Axis not identified | Thin grain | Equivalent, proposed model |
|---|---|---|---|---|---|---|---|
| 1.0 | 40 (11%) [8.5%, 15%] | 8 (20%) | 304 | 6 | 0 | 2 | 350/350 |
| 2.0 | 52 (15%) [12%, 19%] | 1 (1.9%) | 286 | 12 | 0 | 0 | 350/350 |
| 3.0 | 63 (18%) [14%, 22%] | 1 (1.6%) | 270 | 17 | 0 | 1 | 350/350 |
| 4.0 | 68 (19%) [16%, 24%] | 4 (5.9%) | 251 | 28 | 3 | 6 | 345/345 |

## S6. Target tracking before and after relaxation at each guidance strength

The GB energy of each generated structure against its target for the additional experiments of the proposed model at guidance strengths of $w$ = 1.0, 2.0, 3.0, 4.0 is shown in Fig. S1. The top row shows the energies before relaxation (the MLIP single-point energy of the structure immediately after generation, with the same definition of $\gamma$ as Eq. (1)). The bottom row shows the energies after relaxation. The target set is that of the additional experiments (0.32 J/m$^2$ to 2.75 J/m$^2$, Section S5), with five targets inside the training range and two outside it, and 50 structures per target. The 5 structures at $w$ = 4.0 and the target of 0.321 J/m$^2$ that contained elements other than Al (Section S5) are not plotted and are counted as not near-target. The number of near-target structures (within ±0.1 J/m$^2$ of the target) is printed in each panel for each target. The counts before and after relaxation differed by at most 4 structures for any target. Structures whose $\gamma$ exceeded the axis limit of 3.0 J/m$^2$ are drawn at the limit, and their number (8 in total) is printed. The additional experiments were not designed to isolate the effect of guidance strength, and each strength consists of a single generation run. We therefore report these distributions without attributing the differences between panels to guidance strength.

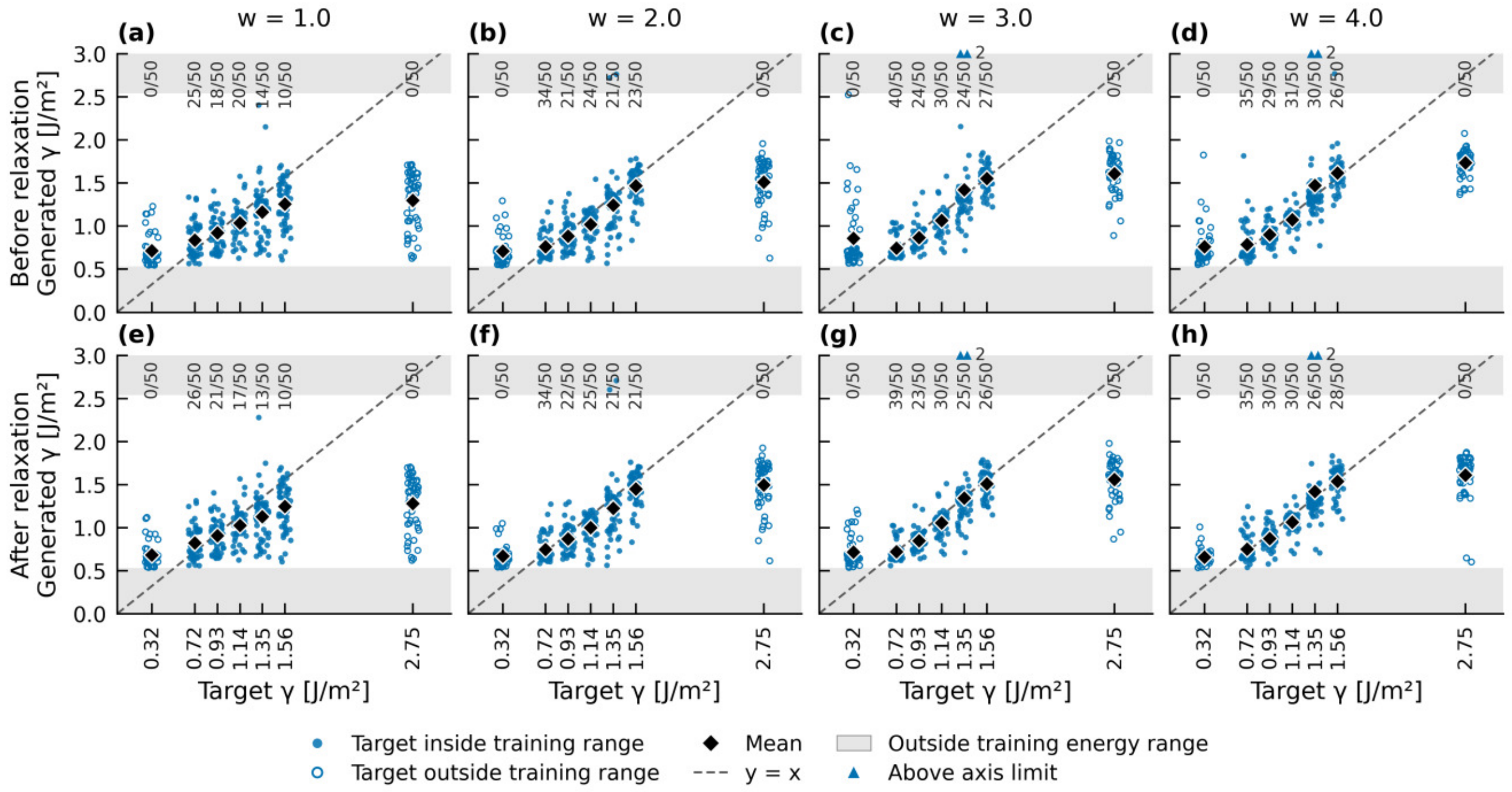

**Fig. S1.** GB energy of the generated structures against the target $\gamma$ before and after relaxation, for the additional experiments of the proposed model at guidance strengths $w$ = 1.0, 2.0, 3.0, 4.0 (columns). All $\gamma$ values are in J/m$^2$. The upper row shows the MLIP single-point $\gamma$ of the structures immediately after generation, and the lower row shows $\gamma$ after relaxation. Panels (a)–(d) are the upper row (before relaxation) and (e)–(h) are the lower row (after relaxation). The columns follow the order of $w$ = 1.0, 2.0, 3.0, 4.0 from the left. In each panel, the 50 structures generated for each of the seven targets are shown as points displaced slightly in the horizontal direction. The mean for each target is shown as a black diamond, and the dashed line is $y = x$. The light gray background marks the outside of the training $\gamma$ range (minimum 0.530 J/m$^2$, maximum 2.537 J/m$^2$). The points of the two targets outside the training range (0.32 J/m$^2$ and 2.75 J/m$^2$) are drawn as open circles. The number above each target is the number of near-target structures (within ±0.1 J/m$^2$ of the target) out of 50. Points with $\gamma$ above 3.0 J/m$^2$ are drawn as triangles at the axis limit together with their number. The 5 structures at $w$ = 4.0 and the target of 0.321 J/m$^2$ that contained elements other than Al are excluded from both rows. The targets differ from those of the main experiment (Fig. 3).

## S7. Exclusion of distorted cells in the unconstrained baseline

Among the 350 structures generated by the unconstrained baseline in the main experiment, 20 were excluded before structural relaxation. For these, at least one of the three cell angles deviated from 90° by more than 1°. The angular deviations of the excluded structures ranged from 1.02° to 1.86°, while the maximum deviation among the remaining 330 structures was 0.79°. The numbers of excluded structures at each target GB energy are summarized in **Table S5**. No structures generated by the proposed model were excluded on the basis of cell-angle deviations. GB equivalence was not assessed for the 20 excluded unconstrained-baseline structures.

**Table S5.** Numbers of unconstrained-baseline structures excluded before relaxation because of cell-angle distortion, at each target GB energy (out of 50 structures per target).

| Target $\gamma$ (J/m$^2$) | Excluded structures (out of 50 each) |
|---|---|
| 0.731 | 4 |
| 1.132 | 1 |
| 1.533 | 0 |

| Target $\gamma$ (J/m$^2$) | Excluded structures (out of 50 each) |
|---|---|
| 1.935 | 2 |
| 2.336 | 1 |
| 0.129 | 11 |
| 2.938 | 1 |

## S8. Grain boundary detection and the equivalence assessment

Only structures in which two GBs were detected in stage 0 proceeded to the equivalence evaluation. The parameters of stage 0 (the bond length, the tilt-axis component, the orientation tolerance, and the layer width) are given with the sensitivity analysis in Section S9. In stage 1, the four candidate symmetry operations were a 180° rotation about the GB normal, a 180° rotation about the tilt axis, a mirror reflection, and an inversion. The mirror plane contains the GB normal and the tilt axis. For each operation, we allowed translation along all three coordinate directions and evaluated the distances between corresponding atoms. In stage 2, we extracted a local window extending ±4.0 Å along the GB normal from the center of each detected boundary. We compared its atomic arrangement with the atoms located within ±8.5 Å of the other boundary. Matching was performed in both directions by exchanging the two GBs, and the larger of the two maximum deviations was used as the evaluation value. The Hungarian assignment (SciPy[45]) minimizes the sum of distances between matched atoms. It does not necessarily minimize the maximum interatomic distance, so its effect on the number of structures classified as equivalent is examined in Section S9. When the grain thickness between the two GBs was less than 17.0 Å (twice the 8.5 Å half-width of the stage 2 window), the local window contained the opposite GB. Stage 2 then could not be judged, and only the result of stage 1 was used. We used about 0.064 Å, obtained by multiplying *stol* = 0.025 of the StructureMatcher by the cube root of the volume per atom of Al (2.54 Å), as a guide, and set the equivalence threshold to 0.064 Å after inspecting the distribution of the maximum deviation in the main experiment (the generation reported in Section 3).

## S9. Sensitivity of grain boundary equivalence assessment

We examined the sensitivity of the GB equivalence assessment to the tolerance. We varied the tolerance over eight values ranging from $1 \times 10^{-4}$ Å to 0.1 Å, including the 0.064 Å used in the main analysis. In stage 0, the local orientation of each atom was determined from in-plane bonds selected among nearest-neighbor bonds within 3.4 Å, with the component along the tilt axis smaller than 0.5 Å. An orientation label was assigned when the local orientation agreed with a reference orientation within 8.0°. The reference orientations were determined from peaks measured using 20 training structures, rather than from the theoretical misorientation of 36.87°. For each 2.0 Å-wide layer along the GB normal, an orientation was accepted when at least two atoms were labeled with that orientation and their fraction was at least 0.6. The number of GBs was then determined from the number of orientation transitions. For the three regenerations of the unconstrained baseline, the numbers of structures classified as containing equivalent GB pairs at the tolerance of 0.064 Å were 3, 2, and 2, respectively. Each regeneration had 350 structures. Even when the tolerance was increased to 0.1 Å, the maximum number classified as equivalent among the four runs (the main experiment and the three regenerations) was 4. In contrast, all 350 structures in each of the two regenerations of the proposed model were classified as equivalent at all eight tolerance values. We repeated the post-relaxation equivalence assessment, replacing the Hungarian assignment of Section S8 with a bottleneck assignment that minimizes the maximum deviation. At a tolerance of 0.064 Å, the results were unchanged. 350 of 350 structures from the proposed model were classified as equivalent. Of the 74 unconstrained-baseline structures containing two GBs, 0 were classified as equivalent. The minimum $d_{\max}$ values among the unconstrained-baseline structures were also unchanged, at 0.193 Å in stage 2 and 0.206 Å in stage 1. Therefore, any tolerance greater than the maximum deviation of $8.5 \times 10^{-6}$ Å for the proposed model and smaller than 0.193 Å gives the same number of equivalent structures for the evaluated structures. In that case, the window would extend to the opposite GB. The distributions of $d_{\max}$ underlying this assessment are shown in **Fig. S**2. The equivalence threshold was set after examining the distribution of geometric distances in the generated structures; it was therefore not fixed by a preregistered procedure.

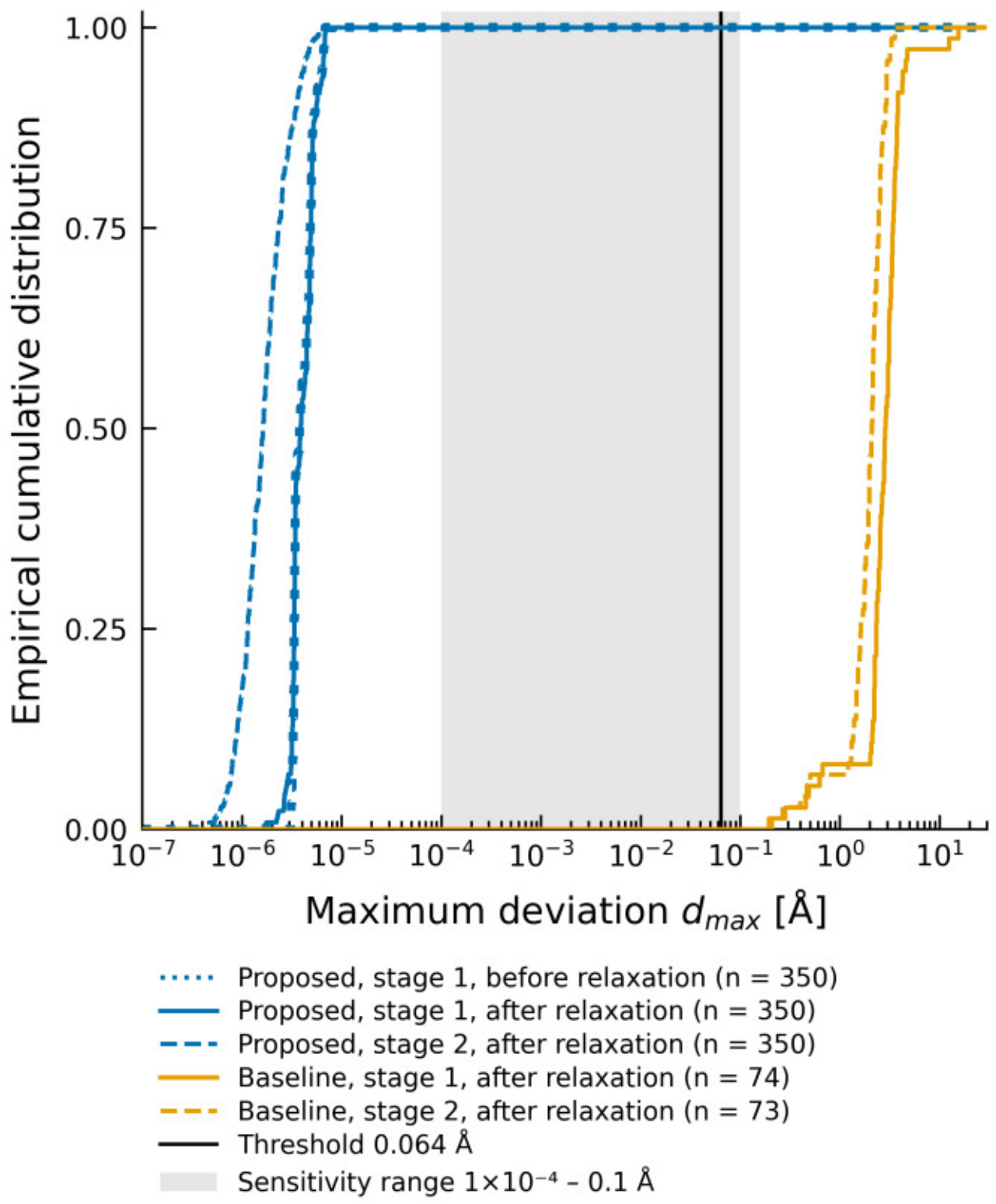

**Fig. S2.** Empirical cumulative distributions of the maximum deviation $d_{\max}$ between corresponding atoms used in the GB equivalence assessment (Section 2.5), for the main experiment. Proposed model: stage 1 (whole-structure matching) before relaxation, and stages 1 and 2 (local window) after relaxation, 350 structures each. Unconstrained baseline: the 74 relaxed structures in which two GBs were detected, with stage 1 and stage 2 after relaxation. One structure with grains too thin for the stage 2 window is omitted from the stage 2 curve. The vertical line marks the equivalence threshold of 0.064 Å, and the gray band marks the range of thresholds from $1 \times 10^{-4}$ Å to 0.1 Å examined in the sensitivity analysis. The largest $d_{\max}$ of the proposed model after relaxation is $8.5 \times 10^{-6}$ Å, and the smallest values for the unconstrained baseline are 0.193 Å (stage 2) and 0.206 Å (stage 1). The horizontal axis is logarithmic.

## S10. Geometric novelty and similarity to training structures

Of the 91 structures generated by the proposed model in the main experiment with relaxed GB energies within ±0.1 J/m$^2$ of their respective targets, 90 did not match any training structure (Fig. S3(a)). Pairwise comparisons among these 90 structures within each target-energy group detected no matching pairs. However, structures generated at different target energies were not compared with one another. These results indicate non-matching with the training structures under the adopted geometric criterion. They do not establish differences at the level of individual GB structural units. Representative near-target, train-unmatched structures (Fig. S4(a)–(c)) and the train-unmatched structure with the lowest $\gamma$ (Fig. S4(d)) are shown together with their nearest training structures.

Each generated structure was compared with training structures having the same number of atoms, selected from the 4,419 training structures. The number of eligible training structures was 225–1,191 depending on the atom count. The main analysis used the StructureMatcher of pymatgen with *stol* = 0.025, *ltol* = 0.2, *angle_tol* = 5°, *primitive_cell=False*, *scale=False*, and *attempt_supercell=False*; elemental species were compared with the default SpeciesComparator, and relaxed structures were used on both sides. The tolerance *stol* = 0.025 corresponds to an absolute tolerance of approximately 0.064 Å when multiplied by the cube root of the volume per Al atom. For the structures that were near-target and train-unmatched, we evaluated the all-atom root-mean-square (RMS) displacement to the nearest training structure. For each generated structure $q$, let $\mathcal{T}_{N_q}$ be the set of relaxed training structures with the same number of atoms $N_q$. We calculated the minimum distance as follows.

$$d_{\mathrm{RMS,min}}(q) = \min_{t \in \mathcal{T}_{N_q}} [d_{\mathrm{RMS}}(q,t)]. \tag{S3}$$

Here, $d_{\mathrm{RMS}}(q,t)$ is the normalized RMS distance returned by *get_rms_dist* of the StructureMatcher. It is converted to Å by multiplying it by the cube root of the volume per atom of the generated structure, $\left(V_q/N_q\right)^{1/3}$. For this distance calculation, *stol* = 0.6 was used to obtain the geometric correspondence. The other settings, including *ltol*, *angle_tol*, and cell scaling, followed the matching conditions described above. *get_rms_dist* selects the atom correspondence that minimizes the RMS. The calculation uses the displacements of the whole cell and all atoms, taking the periodic boundary conditions and the atom correspondence into account. The translational component is removed after the atoms are matched. It is not an RMS restricted to the atoms near the GBs. The match judgment against the training data (*stol* = 0.025) is based on the maximum site displacement of corresponding atoms. It is therefore a different quantity from $d_{\mathrm{RMS,min}}$.

To assess sensitivity to this parameter, we varied *stol* across four values: 0.015, 0.025, 0.040, and 0.060. The corresponding absolute tolerance ranges, calculated across the lattices of the 4,419 training structures, were 0.038–0.039 Å, 0.063–0.066 Å, 0.101–0.105 Å, and 0.151–0.157 Å, respectively. Across all four levels, these tolerances spanned approximately 0.038 Å to 0.157 Å. The numbers of near-target structures that did not match any training structure were 91, 90, 84, and 66, respectively. These counts represent train-unmatched structures. They should be distinguished from the numbers of groups obtained by pairwise comparisons among generated structures within each target-energy group. Separating the near-target structures by whether the generated $dx$ lay inside or outside the training range, the numbers of train-unmatched structures were 48, 48, 44, and 30 of 48 inside the range and 43, 42, 40, and 36 of 43 outside it, for *stol* = 0.015, 0.025, 0.040, and 0.060, respectively. We further examined the discrimination provided by StructureMatcher using pairs of training structures whose $dx$ values differed by one step of 0.15 Å. At *stol* = 0.025, 166 such pairs (5.0%) were classified as matching. As *stol* increased through the four values, the numbers of groups obtained by comparisons within each target-energy group were 91, 90, 84, and 65 after relaxation. Before relaxation, they were 97, 96, 91, and 73. Meanwhile, the numbers of adjacent-$dx$ training-structure pairs classified as matching increased as 0, 166, 257, and 321. As additional checks of the matching procedure, we applied identity-preserving transformations (translation, atom-order permutation, rotation, and others) to training structures, and 20/20 were classified as matching. When we applied random displacements exceeding the tolerance, 20/20 were classified as non-matching.

We did not assign a binary classification to whether a generated structure occupied a position between structures represented in the training data. For the 90 train-unmatched structures, the median $d_{\mathrm{RMS,min}}$ was 0.165 Å (**Fig. S**3(b)). This was smaller than both the median for training-structure pairs with adjacent $dx$ values (0.455 Å) and the median for the test-split structures (0.184 Å). Here the median for the test-split structures is the median of $d_{\mathrm{RMS,min}}$ to the nearest train structure. It is taken over 120 structures drawn at random from the 552 test-split structures, in proportion to the number of structures at each atom count. The test-split structures come from templates not included in train. Among the 90 structures, 81 had a smaller distance to the training data than the median for adjacent-$dx$ pairs. In contrast, 40 had a larger distance than the median for those test-split structures. Finally, we examined the 10 structures with the lowest GB energies among the generated structures from the proposed model's main experiment that were train-unmatched and converged during relaxation. Their minimum energy was 0.535 J/m$^2$, compared with the training-data minimum of 0.530 J/m$^2$. Their minimum all-atom RMS displacements to the nearest training structures ranged from 0.075 Å to 0.160 Å, with a median of 0.105 Å. Visual inspection suggested that these structures contained kite-type structural units[5] similar to those observed in the training data.

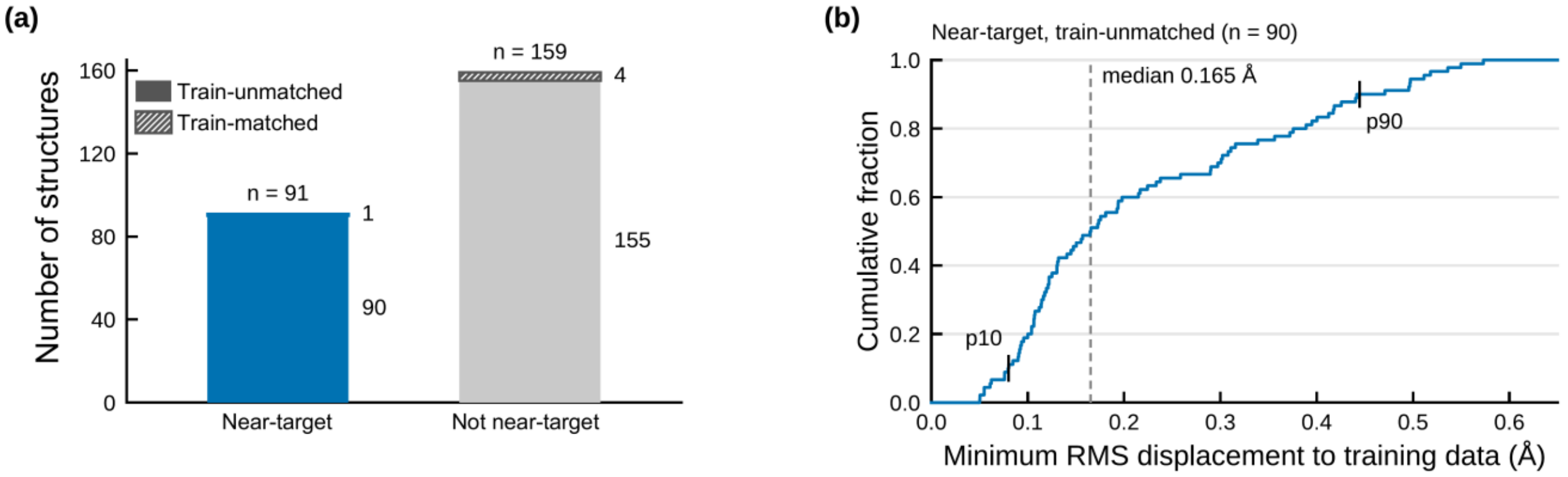

**Fig. S3.** Comparison of the generated structures with the training data. (a) Classification of the 250 structures generated by the proposed model ($w = 2.0$) for the five targets inside the training range (50 for each target). The left bar shows the 91 near-target structures, whose relaxed $\gamma$ was within ±0.1 J/m$^2$ of the target, and the right bar shows the remaining 159 structures. Each bar is divided by whether the structure matches a training structure with the same number of atoms (StructureMatcher, *stol* = 0.025): train-unmatched (solid) and train-matched (hatched). (b) Empirical cumulative distribution of the minimum all-atom RMS displacement to the nearest training structure, for the 90 structures that are near the target and do not match the training data. The median is a dashed line, and the 10th and 90th percentiles are short vertical ticks. Targets outside the training range and the unconstrained baseline are not included.

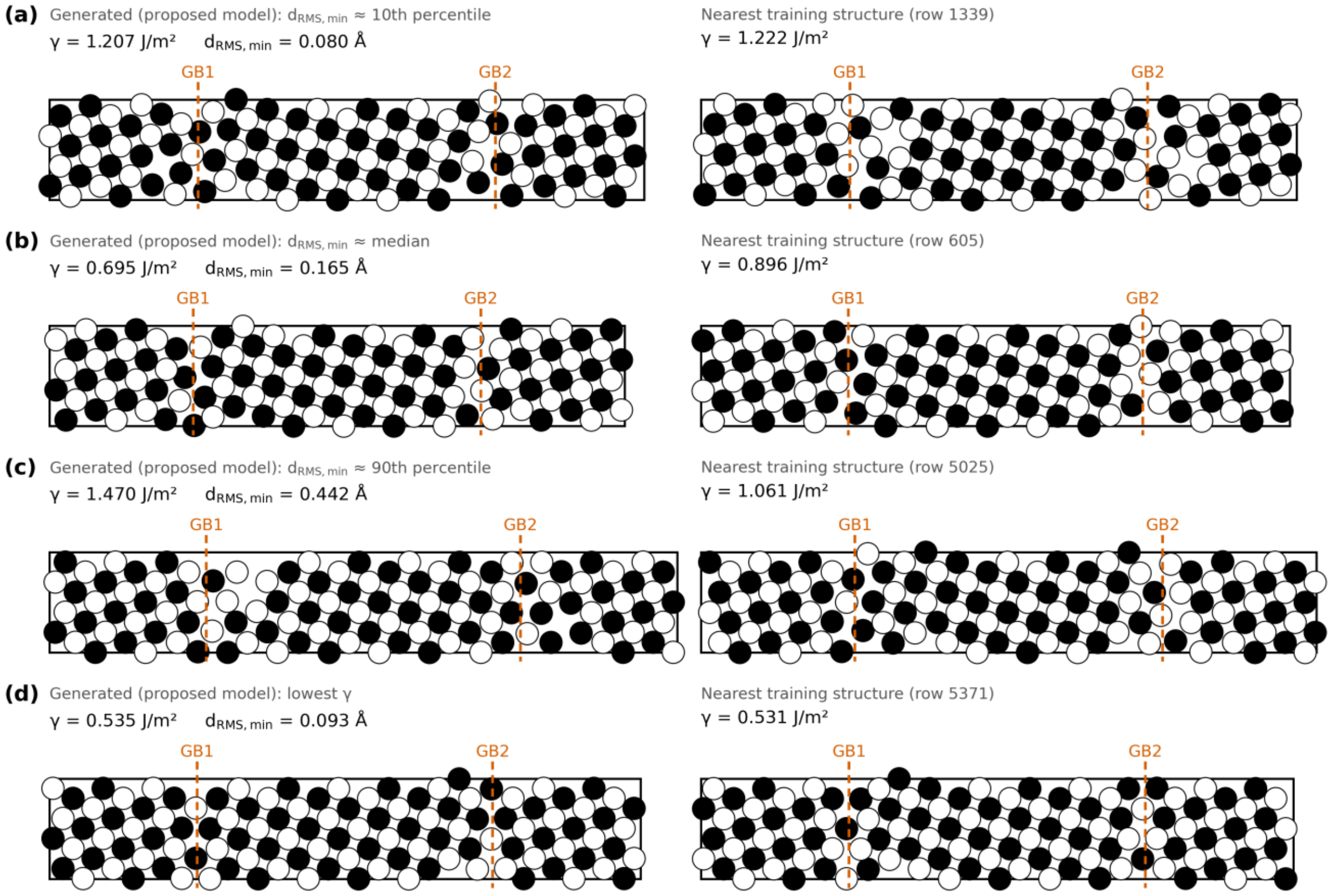

**Fig. S4.** Representative generated structures of the proposed model (main experiment, left column) and their nearest training structures (right column), viewed along the [001] tilt axis. Filled and open circles are the two atomic layers along the projection direction, the black rectangle is the periodic cell, and the orange dashed lines mark the detected GB positions (GB1 and GB2). The structures are translated periodically so that the two GBs sit symmetrically about the cell center, as in Fig. 2. (a)–(c) Near-target, train-unmatched structures. Their minimum all-atom RMS displacements to the nearest training structure, $d_{\mathrm{RMS,min}}$, are closest to the 10th percentile (0.080 Å), the median (0.165 Å), and the 90th percentile (0.445 Å), respectively, of the 90 such structures (Fig. S3(b)). (d) The train-unmatched structure with the lowest relaxed $\gamma$ (0.535 J/m$^2$, Section S10), which was generated for the target below the training range and is not near-target. The nearest training structure is the one that minimizes $d_{\mathrm{RMS}}$ among the relaxed training structures with the same number of atoms (Section S10). Its $\gamma$ is given above each panel. The values of $\gamma$ and $d_{\mathrm{RMS,min}}$ are printed above each generated structure. In each panel heading, row is the row number in the training dataset.

## S11. Unconditional generation and training-data baseline

To assess the contribution of property conditioning to target tracking, we used two controls. One was unconditional generation with the same random seed. The other was the near-target fraction expected from random sampling of the training dataset. For the first control, we compared structures generated without conditioning ($w = 0$) with those generated by the proposed model at the corresponding targets. We used 50 structures per target and the same initial seed. For the second control, we calculated the fraction of training structures whose energies fell within ±0.1 J/m$^2$ of each target. Unrounded values were used and the boundary was included. This fraction corresponds to the probability of obtaining a near-target structure by randomly selecting one structure from the training dataset. For the three targets with relatively many nearby training structures, the training-data base rates were 16%, 22%, and 9.7%. For unconditional generation ($w = 0$), the relaxed GB energies ranged from 0.540 J/m$^2$ to 1.649 J/m$^2$, and all structural relaxations converged. This energy range did not extend to the target values of 2.336 J/m$^2$ and 2.938 J/m$^2$. When the 50 unconditionally generated structures were paired with each target using the same seed, the numbers falling within ±0.1 J/m$^2$ of 0.731 J/m$^2$, 1.132 J/m$^2$, 1.533 J/m$^2$, and 1.935 J/m$^2$ were 12, 7, 0, and 0, respectively. For the remaining targets of 2.336 J/m$^2$, 0.129 J/m$^2$, and 2.938 J/m$^2$, the number was 0. **Table S6** compares the numbers of near-target structures obtained with the proposed model ($w = 2.0$) and unconditional generation ($w = 0$) using the same seed. The targets are listed in ascending order of GB energy, and near-target structures are defined by a relaxed GB energy within ±0.1 J/m$^2$ of the target. For the same two sets of structures, **Fig. S5** shows the GB energies immediately after generation and after relaxation against the target.

**Table S6.** Numbers of near-target structures obtained with the proposed model in three independent generation series (series 1 is the main experiment; series 2 and 3 are regenerations with different random seeds) and with unconditional generation using the same seed as series 1, listed in ascending order of target GB energy, together with the number of training structures near each target and the training-data base rate. Near-target structures are defined by a relaxed GB energy within ±0.1 J/m$^2$ of the target; counts are out of 50 structures per target. The ratio to the base rate is the near-target fraction (near-target count/50) divided by the training-data base rate.

| **Target $\gamma$ (J/m$^2$)** | **Training structures within ±0.1 J/m$^2$** | **Proposed model ($w$ = 2.0)** | **Proposed model, series 2** | **Proposed model, series 3** | **Unconditional ($w$ = 0, same seed)** | **Training-data base rate (%)** |
|---|---|---|---|---|---|---|
| 0.129 | 0 | 0/50 | 0/50 | 0/50 | 0/50 | 0.0 |
| 0.731 | 714 | 35/50 | 33/50 | 35/50 | 12/50 | 16 |
| 1.132 | 985 | 22/50 | 24/50 | 27/50 | 7/50 | 22 |
| 1.533 | 427 | 25/50 | 22/50 | 22/50 | 0/50 | 9.7 |
| 1.935 | 85 | 9/50 | 12/50 | 15/50 | 0/50 | 1.9 |
| 2.336 | 0 | 0/50 | 0/50 | 0/50 | 0/50 | 0.0 |
| 2.938 | 0 | 0/50 | 0/50 | 0/50 | 0/50 | 0.0 |

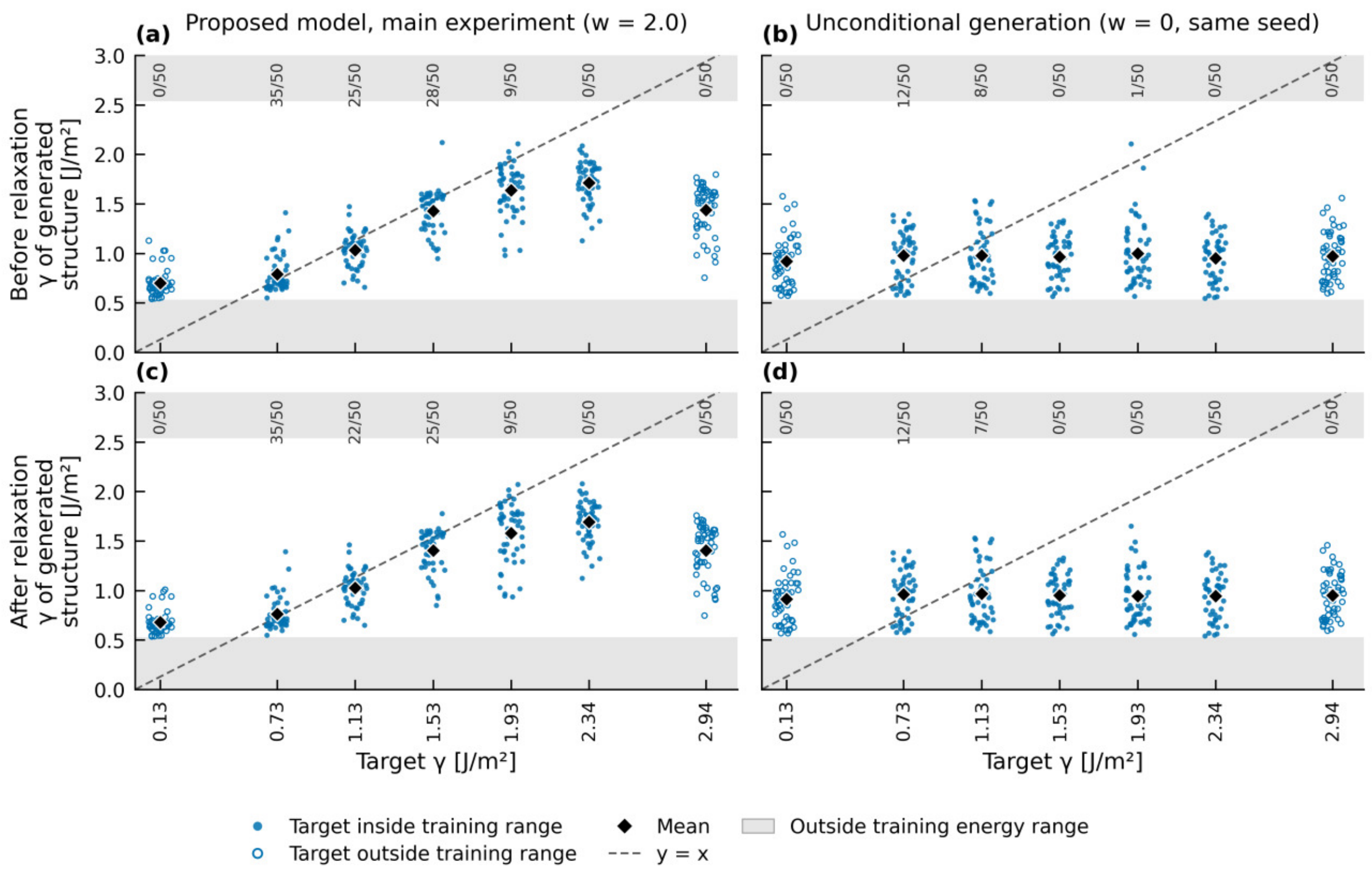

**Fig. S5.** GB energies of the generated structures against the target $\gamma$, before and after relaxation. The left column is the main experiment of the proposed model ($w$ = 2.0). The right column is unconditional generation with the same seed ($w$ = 0). All $\gamma$ values are in J/m$^2$. The top row, (a) and (b), shows the MLIP single-point $\gamma$ of the structures immediately after generation (with the same definition of $\gamma$ as Eq. (1)). The bottom row, (c) and (d), shows $\gamma$ after relaxation, and (c) contains the same 350 structures as Fig. 3. The targets on the horizontal axis are the seven targets of the main experiment. For $w$ = 0 they are nominal, because the condition does not affect the generation. In each panel, the 50 structures per target are drawn as points with a small horizontal offset (filled for targets inside the training range, open for targets outside it). The black diamonds are the per-target means, and the dashed line is $y = x$. The light gray background marks the region outside the training-data $\gamma$ range (0.530 J/m$^2$ to 2.537 J/m$^2$). The label $k$/50 above each target gives the number of near-target structures (within ±0.1 J/m$^2$). The near-target counts before and after relaxation differed by at most 3 structures for any column and target.

## S12. Application to the BCC Fe Σ5(310)[001] symmetric tilt grain boundary

To check whether the procedure transfers to a different crystal structure, we applied it to the body-centered cubic (BCC) Fe Σ5(310)[001] symmetric tilt GB with the same model architecture and training settings as for Al. Only system-specific constants were changed: the paths of the training data, the distribution of the number of atoms, the mean and variance of the rigid translation $dx$, the atomic number of the host element, and the file of Fe-specific cell constants. The training dataset was built following the procedure of Section S1 from 31 templates; after duplicate removal it contained 3,720 relaxed structures. Templates whose relaxed structures coincided were assigned to the same class, and the split was made by class into 3,105, 289, and 326 structures for training, validation, and test. Each cell contained 36 fixed bulk atoms and 72 to 80 atoms in total.

Because MACE-MH-1 does not predict magnetic moments, we compared MLIP and spin-polarized density functional theory (DFT) GB energies for 12 GB structures (six without and six with removed atom pairs, selected by their MLIP energies) before generating structures. The DFT calculations used the Perdew–Burke–Ernzerhof (PBE) functional, a cutoff of 520 eV, Methfessel–Paxton smearing of 0.10 eV, and ferromagnetic initial moments of 2.5 $\mu_B$ per atom. The MLIP and PBE lattice constants of bulk BCC Fe differed by 0.39% in relative terms. For the 12 structures, $\rho$ between the MLIP and DFT energies was 0.986, and the sign of the energy difference between two structures agreed for 59 of the 60 pairs whose DFT energies differed by at least 0.10 J/m$^2$. The MLIP energies were lower than the DFT energies, with a median MLIP − DFT difference of −0.165 J/m$^2$. These checks concern GB structures built by the procedure used for the training data, before the training data were generated; DFT calculations on the generated Fe structures were not performed.

The checkpoint with the lowest validation loss was used, as for Al (**Table S2**); it was epoch 189 of 200, with a validation loss of 0.629. At guidance strength $w$ = 2.0, 50 structures were generated for each of five targets from 1.811 J/m$^2$ to 4.963 J/m$^2$. The targets were chosen by the rule used for Al (Section S5): they were placed at 10%, 30%, 50%, 70%, 90% of the width of the training energy range above its lower bound.

All 250 generated structures were valid: they consisted only of Fe atoms and were relaxed with the MLIP to a maximum force of at most 0.02 eV/Å within 500 steps, with finite energy and $\gamma$. Of these, 250 were judged equivalent, that is, the best residual over four candidate symmetry operations was at most 0.064 Å. A generated structure and a training structure selected by a rule fixed beforehand are shown in **Fig.** S6. We calculated $\rho$ between the target and the MLIP $\gamma$ for the five targets pooled, with 95% bootstrap intervals from resampling stratified by target (2,000 draws). It was 0.808 (interval 0.765 to 0.847) before relaxation and 0.766 (0.711 to 0.816) after relaxation. After relaxation, 24, 7, 4, 5, and 3 of the 50 structures at the five targets, in ascending order, were within ±0.1 J/m$^2$ of their targets (43 of 250 in total). Before relaxation, the median $\gamma$ differed from the target by 0.002 J/m$^2$ at the lowest target and was lower than the target at each of the other four, by as much as 0.647 J/m$^2$ at the highest. Only 5 training structures lay within ±0.1 J/m$^2$ of the highest target, fewer than the 62 (2% of the 3,105 training structures) that we set as the criterion for a target to be regarded as supported by the training data. The distributions and the values are given in Fig. S7 and Table S7.

**Table S7.** Summary of the BCC Fe Σ5(310)[001] check, for the five targets inside the training range ($w$ = 2.0, 50 structures per target, 250 structures in total). $\rho$ is the Spearman correlation between the target and the MLIP $\gamma$ for the five targets pooled, with the 95% bootstrap interval stratified by target in brackets. The counts per target are the numbers of structures within ±0.1 J/m$^2$ of the target after relaxation, out of 50, for the targets in ascending order. A structure was equivalent if the best residual over four candidate symmetry operations was at most 0.064 Å, and the denominator was the valid structures.

| **Item** | **Value** |
|---|---|
| Structures in the dataset after duplicate removal | 3,720 |
| Targets (J/m$^2$) | 1.811 to 4.963 |
| Valid structures | 250 |
| Equivalent structures | 250 |
| $\rho$ before relaxation (95% bootstrap interval) | 0.808 (0.765, 0.847) |
| $\rho$ after relaxation (95% bootstrap interval) | 0.766 (0.711, 0.816) |
| Near-target structures at each target | 24, 7, 4, 5, 3 |

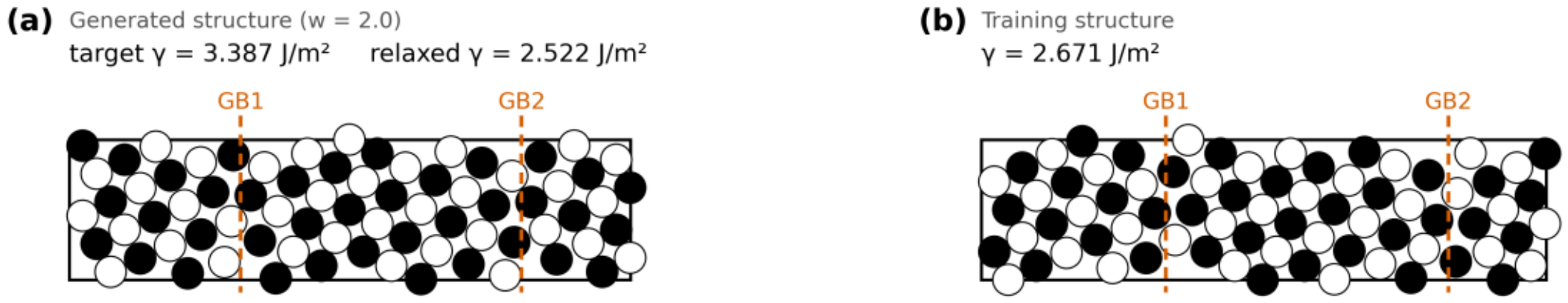

**Fig. S6.** Representative structures of BCC Fe Σ5(310)[001], viewed along the [001] tilt axis: (a) a generated structure ($w$ = 2.0) and (b) a training structure. Both were selected by a rule fixed before the structures were examined. The generated structure is the one whose relaxed $\gamma$ is closest to the median (2.526 J/m$^2$) of the 250 structures that were valid and equivalent among those generated for the five targets. The training structure is the one whose $\gamma$ is closest to the median (2.671 J/m$^2$) of the 3,105 structures of the training split. When two structures were equally close, the one with the smaller index was chosen. The rule selects by $\gamma$ alone, so the generated structure is not a near-target example: its target was 3.387 J/m$^2$, whereas its relaxed $\gamma$ was 2.522 J/m$^2$. Filled and open circles are the two atomic layers along the projection direction, and the black rectangle is the periodic cell. The orange dashed lines mark the GB centers (GB1 and GB2), taken as the mean position along the boundary normal of atoms with a BCC centrosymmetry parameter above 1.0 Å$^2$. The structures are translated periodically so that the two boundaries sit near the cell center, as in Fig. S4. Above each panel are the target $\gamma$ and the relaxed $\gamma$ of the generated structure in (a) and the $\gamma$ of the training structure in (b), in J/m$^2$.

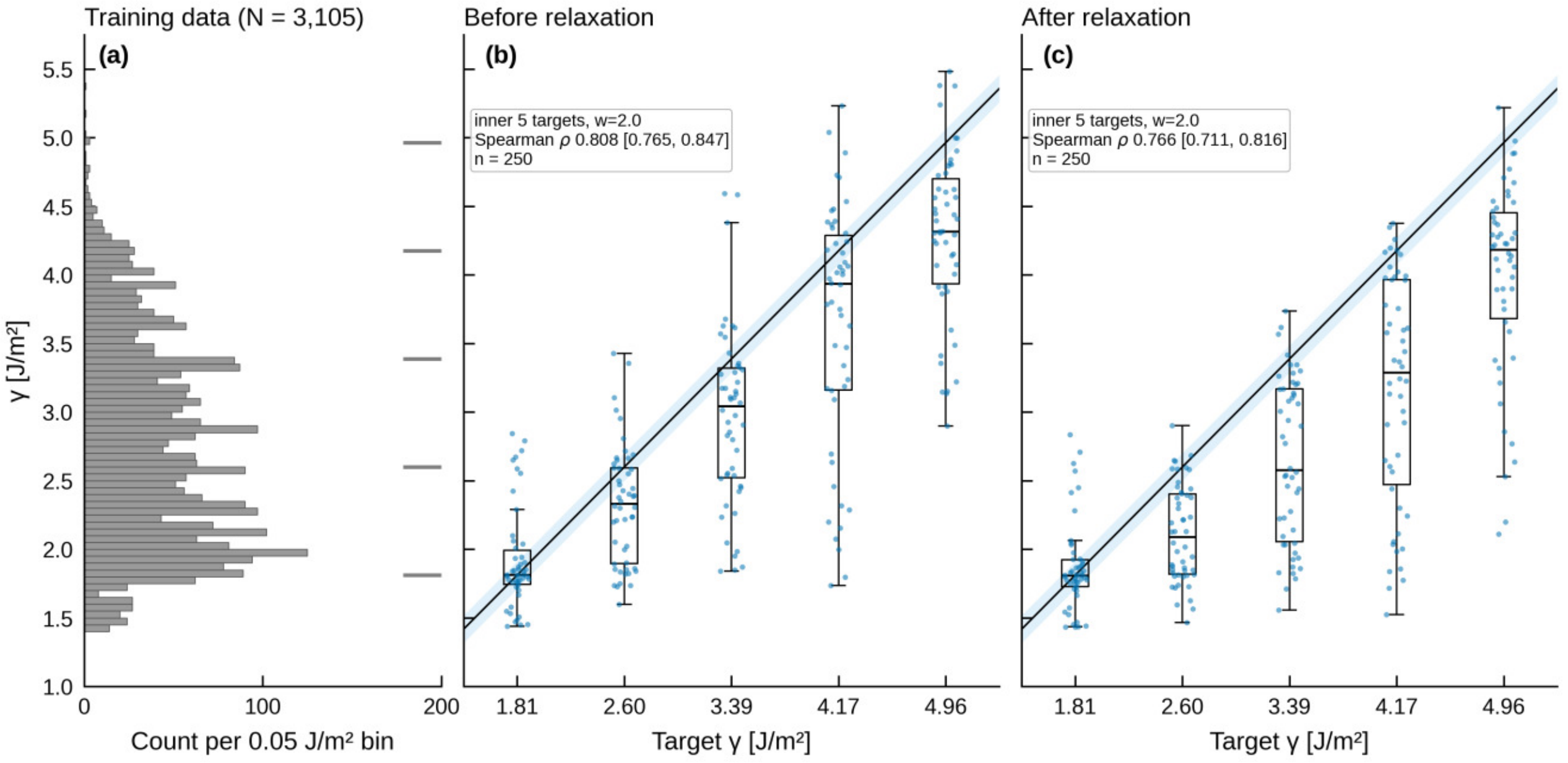

**Fig. S7.** Tracking of the target $\gamma$ for BCC Fe Σ5(310)[001]. The vertical axis is the $\gamma$ common to all panels, and all $\gamma$ values are in J/m$^2$. (a) Horizontal histogram of the $\gamma$ of the 3,105 structures of the training split of the Fe data (bin width 0.05 J/m$^2$); short horizontal lines mark the five target $\gamma$ values. (b) and (c) show the structures generated at guidance strength $w$ = 2.0 for the five targets, which lie inside the training range (50 structures per target, 250 in total), against the target $\gamma$: (b) is the MLIP single-point $\gamma$ of the structures immediately after generation, and (c) is $\gamma$ after relaxation. In (b) and (c), the black line is $y = x$, and the light blue band is ±0.1 J/m$^2$ around it. The points are the individual structures, displaced slightly in the horizontal direction. Each box spans the quartiles with a line at the median, and the whiskers extend to the most extreme points within 1.5 times the interquartile range. The $\rho$ in (b) and (c) is for the 250 structures pooled over the targets, and the interval in brackets is the 95% bootstrap interval stratified by target.

This check was more limited than the Al evaluation. The unconstrained baseline, DFT calculations on generated structures, and the novelty assessment were not repeated; equivalence was assessed only by the best residual over four candidate symmetry operations, with the threshold of Section S8 and without the three-stage procedure of that section; the evaluation covered only the five targets inside the training range; the validation set consisted of a single class of templates; a single training seed was used; and the acceptance thresholds were set at levels that the Al result satisfies. The result shows that the procedure can be applied to a BCC boundary with only system-specific changes; it does not show that its performance is maintained across crystal structures.

## S13. DFT single-point validation

The DFT single-point calculations were performed on 30 structures drawn at random from the 91 near-target structures of the proposed model in the main experiment ($w$ = 2.0), stratified by target in proportion to the number of near-target structures at each target (12, 7, 8, and 3 structures for the targets 0.731 J/m$^2$, 1.132 J/m$^2$, 1.533 J/m$^2$, and 1.935 J/m$^2$; no near-target structure was obtained at 2.336 J/m$^2$). The structures were taken after the same fixed-bulk, fixed-cell relaxation used for the training labels, and no DFT relaxation was performed. The calculations used the PBE exchange–correlation functional[46], the projector augmented-wave method[47,48], VASP[49] 6.4.2, an energy cutoff of 520 eV, an electronic convergence criterion of $1 \times 10^{-6}$ eV, and a Γ-centered $1 \times 7 \times 3$ $k$-point mesh. The bulk reference was a perfect FCC Al cell with the same in-plane lattice as the GB structures, which is common to all 30 structures, so a single bulk calculation was used. All calculations converged. The DFT GB energies were lower than the MLIP values for all but 1 of the 30 structures, with a median DFT − MLIP difference of −0.128 J/m$^2$; the median difference grew in magnitude with the target, from −0.096 J/m$^2$ at 0.731 J/m$^2$ to −0.214 J/m$^2$ at 1.935 J/m$^2$ (**Table S**8). The mean absolute error was 0.127 J/m$^2$ before and 0.041 J/m$^2$ after subtracting the common median offset. Over the 30 structures, $\rho$ between the DFT and MLIP energies was 0.988, with a 95% bootstrap interval of 0.955 to 0.996 (resampling the structures with replacement, 10,000 draws). The correspondence of the individual values is shown in **Fig. S**8.

**Table S8.** Median difference between the DFT and MLIP GB energies (DFT − MLIP) for the randomly drawn structures, by target.

| Target (J/m²) | $n$ | Median DFT − MLIP (J/m²) |
|---|---|---|
| 0.731 | 12 | −0.096 |
| 1.132 | 7 | −0.128 |
| 1.533 | 8 | −0.170 |
| 1.935 | 3 | −0.214 |
| All | 30 | −0.128 |

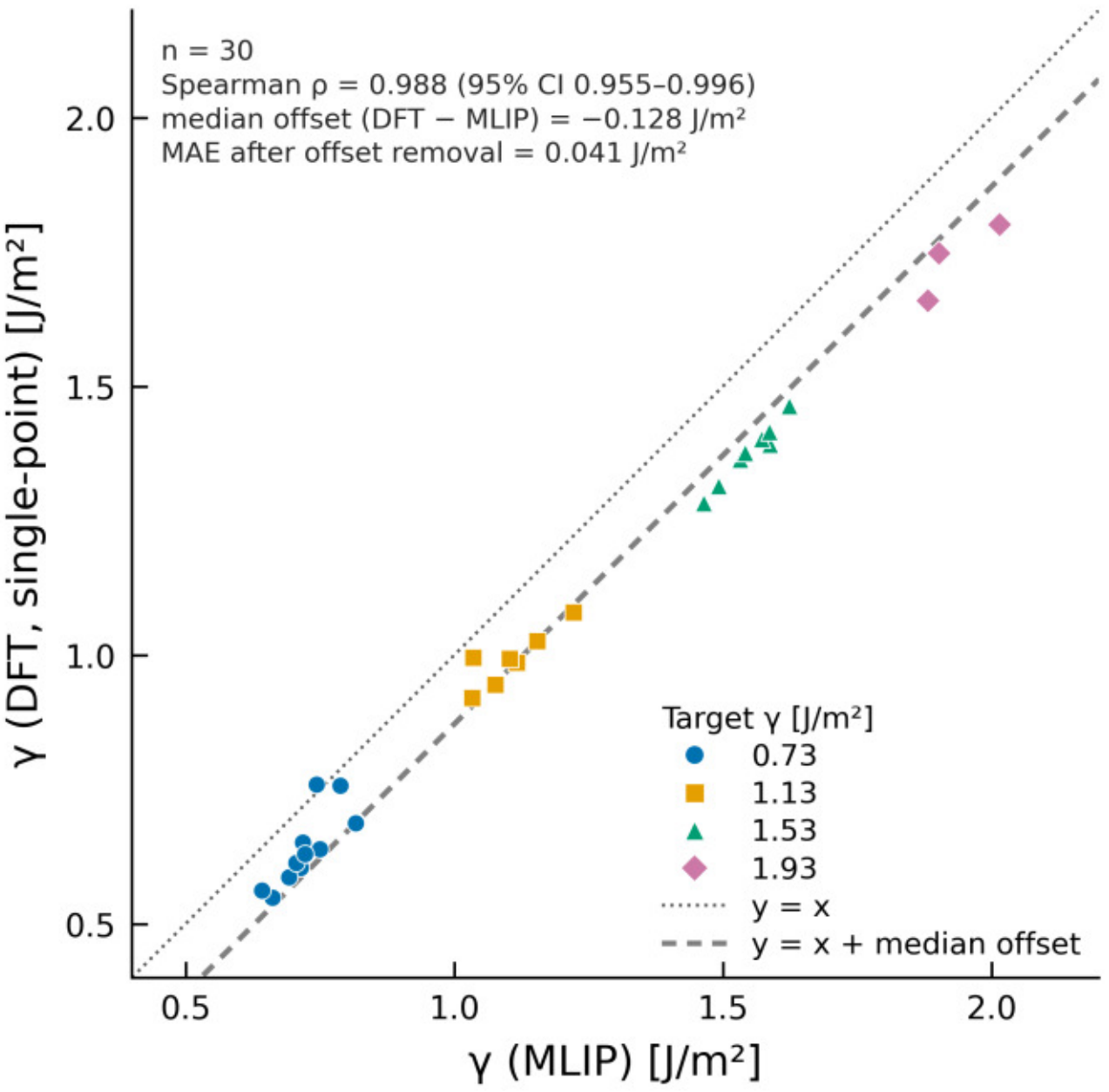

**Fig. S8.** Correspondence between the DFT single-point and MLIP GB energies for 30 structures drawn at random, stratified by target, from the 91 near-target structures of the proposed model ($w$ = 2.0); marker color and shape give the target $\gamma$. The MLIP value is the $\gamma$ after the same fixed-bulk, fixed-cell relaxation used for the training labels, and the DFT value is a single-point calculation on that relaxed structure. All $\gamma$ values are in J/m². The MLIP ordering is preserved in DFT ($\rho$ = 0.988, 95% bootstrap interval 0.955 to 0.996), while DFT is systematically lower than MLIP, with a median DFT − MLIP difference of −0.128 J/m² (the dashed line is $y = x$ shifted by this difference; the dotted line is $y = x$). The difference grows in magnitude with the target (target-wise medians −0.096 J/m² to −0.214 J/m²), and the mean absolute deviation after removing the median offset is 0.041 J/m².

## S14. Distribution of the generated rigid translation $dx$

The rigid translation $dx$ of Bulk B relative to Bulk A along the GB normal, as generated by the proposed model in the main experiment for each of the seven targets, is shown in Fig. S9. It also shows the range of $dx$ used during training (−0.75 Å to +0.60 Å in steps of 0.15 Å). Of the 350 structures, 207 had $dx$ outside this range (Section 4.5). Among the 91 near-target structures at the five targets within the training range, 48 had $dx$ inside the range and 43 outside it (Section 3.3). The number of structures with $dx$ outside the training range is printed for each target. The figure describes the distribution only. The physical validity of the structures with $dx$ outside the training range was not examined (Section 4.5). Among the 250 structures generated for the five targets inside the training range, 48 of the 104 structures with $dx$ inside the training range and 43 of the 146 structures with $dx$ outside it were near-target; the target-stratified Mantel–Haenszel odds ratio was 0.59 (Robins–Breslow–Greenland 95% confidence interval 0.29 to 1.22; within-target permutation test, $p$ = 0.22), so the data do not resolve whether $dx$ outside the range lowers the near-target fraction. Within each target, the generated $dx$ and the number of atoms were correlated ($\rho$ = 0.623, permutation $p \approx 2 \times 10^{-4}$), whereas in the training data they were not (0.014). The distribution of $N$ in the generated structures followed that of the training data (**Table S9**).

**Table S9.** Distribution of the number of atoms $N$ in the training structures and in the structures generated by the proposed model (main experiment).

| $N$ | Training (%) | Generated (%) |
|---|---|---|
| 112 | 27 | 25 |
| 114 | 26 | 26 |
| 116 | 23 | 23 |
| 118 | 19 | 23 |
| 120 | 5.1 | 3.1 |

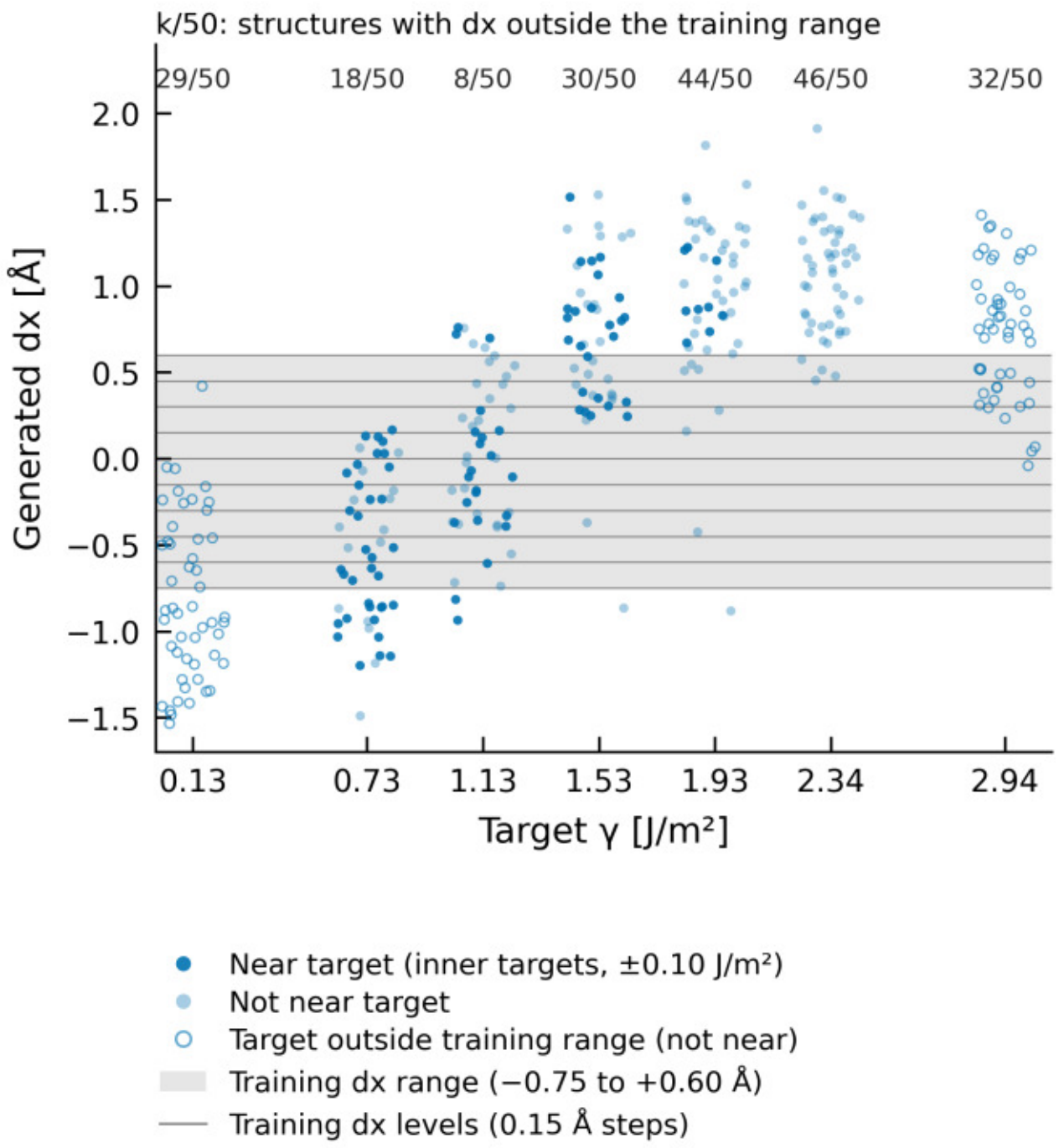

**Fig. S9.** Rigid translation $dx$ generated by the proposed model in the main experiment ($w$ = 2.0), plotted against the target $\gamma$. Each target has 50 structures, displaced slightly in the horizontal direction. Filled dark points are near-target structures (relaxed $\gamma$ within ±0.1 J/m$^2$ of the target) at the targets inside the training range, and light points are the other structures. Open circles are the structures generated for the two targets outside the training range (none of which is near-target). The gray band marks the range of $dx$ used during training (−0.75 Å to +0.60 Å), and the thin horizontal lines mark its 10 levels in steps of 0.15 Å. The number above each target is the number of structures whose $dx$ lies outside the training range, out of 50.